\documentclass[sigconf]{acmart}
\usepackage{booktabs}
\usepackage{amsmath,graphicx,braket,subfigure}
\usepackage{multirow}
\usepackage{balance}

\AtBeginDocument{%
  \providecommand\BibTeX{{%
    \normalfont B\kern-0.5em{\scshape i\kern-0.25em b}\kern-0.8em\TeX}}}

\copyrightyear{2022}
\acmYear{2022}
\setcopyright{acmcopyright}\acmConference[CHI '22]{CHI Conference on Human Factors in Computing Systems}{April 29-May 5, 2022}{New Orleans, LA, USA}
\acmBooktitle{CHI Conference on Human Factors in Computing Systems (CHI '22), April 29-May 5, 2022, New Orleans, LA, USA}
\acmPrice{15.00}
\acmDOI{10.1145/3491102.3517471}
\acmISBN{978-1-4503-9157-3/22/04}

\begin{document}

\title{Impacts of Personal Characteristics on User Trust in Conversational Recommender Systems}

\author{Wanling Cai}
\affiliation{%
  \institution{Department of Computer Science, Hong Kong Baptist University}
  \city{Hong Kong}
  \country{China}
}
\email{cswlcai@comp.hkbu.edu.hk}

\author{Yucheng Jin}
\affiliation{%
  \institution{Department of Computer Science, Hong Kong Baptist University}
  \city{Hong Kong}
  \country{China}
}
\email{yuchengjin@hkbu.edu.hk}

\author{Li Chen}
\affiliation{%
  \institution{Department of Computer Science, Hong Kong Baptist University}
  \city{Hong Kong}
  \country{China}
}
\email{lichen@comp.hkbu.edu.hk}

\renewcommand{\shortauthors}{Cai, et al.}


\begin{abstract}

Conversational recommender systems (CRSs) imitate human advisors to assist users in finding items through conversations and have recently gained increasing attention in domains such as media and e-commerce. Like in human communication, building trust in human-agent communication is essential given its significant influence on user behavior. However, inspiring user trust in CRSs with a ``one-size-fits-all'' design is difficult, as individual users may have their own expectations for conversational interactions (e.g., who, \textit{user} or \textit{system}, takes the initiative), which are potentially related to their personal characteristics. In this study, we investigated the impacts of three personal characteristics, namely \textit{personality traits}, \textit{trust propensity}, and \textit{domain knowledge}, on user trust in two types of text-based CRSs, i.e., user-initiative and mixed-initiative. Our between-subjects user study (N=148) revealed that users' \textit{trust propensity} and \textit{domain knowledge} positively influenced their trust in CRSs, and that users with high \textit{conscientiousness} tended to trust the mixed-initiative system.


 
\end{abstract}



\maketitle

\section{Introduction}
Conversational recommender systems (CRSs) imitate human advisors to assist users in finding desired items through multi-turn conversations and have been attracting increasing attention in recent years for developing task-oriented chatbots~\cite{Christakopoulou:2016,Cai:UMAP2020,jannach2020survey}. Some commercial chatbots have been built on Facebook Messenger or Amazon Alexa for recommending items (e.g., songs, movies, and products)~\cite{sun2018conversational}. Unlike traditional recommender systems that mainly present one-shot recommendations (e.g., a ranked list of items) to users~\cite{Ricci:2010:RSH:1941884}, CRSs can support \textit{mixed-initiative} (combining both \textit{user-initiative} and \textit{system-initiative}~\cite{Allen:1999}) interactions between users and the system~\cite{Cai:UMAP2020}. In such a system, users can not only actively inform the system of their preferences (e.g., ``\textit{I want relaxing music.}''), but also accept proactive suggestions from the system (e.g., ``\textit{Do you want to try some piano music?}'')~\cite{jannach2020survey}. Recent works have shown that mixed-initiative CRSs can help users better control the recommendation~\cite{Jin:CIKM2019} and facilitate their exploration~\cite{Cai:IUI2021}.

However, few studies have investigated the influence of the conversational interaction -- particularly the initiative strategy (i.e., who, \textit{user} or \textit{system}, takes the initiative in the conversation) -- on user trust in CRSs~\cite{jannach2020survey}. Given that user trust plays a vital role in users' willingness to accept recommendations~\cite{Shlomo:IUI2017} and adopt a given system~\cite{benbasat2005trust,chen2005trust}, which can be inherently affected by users' personal characteristics (such as personality traits) \cite{Bart:RecSys2011,Millecamp:UMAP2018}, this work aims to identify whether and how user characteristics and system conversation design factors affect user trust in text-based CRSs. Our findings will be useful for optimizing the design of CRSs to be more trustworthy for individual users, which may potentially maximize the benefit of CRSs.

Our work is theoretically driven by the three-layered trust model proposed by Hoff and Bashir~\cite{hoff2015trust}, which suggests that user trust in a computer system can be influenced by three types of factors: \textbf{user-related factors},  \textbf{system-related factors}, and \textbf{context-related factors}. Among user-related factors, inspired by previous works~\cite{cho2016effect,zhou2020effects,Bart:RecSys2011}, we considered three personal characteristics: (1) \textit{personality traits}, which refer to enduring characteristics related to people's thinking, feeling, and behaving, and have been shown to influence user trust in both human-human and human-machine relationships~\cite{cho2016effect, zhou2020effects};  (2) \textit{trust propensity}, which can be defined as the user's general tendency to trust others, and has been demonstrated to impact user trust in traditional recommender systems~\cite{chen2005trust,Bart:RecSys2011}; and (3) \textit{domain knowledge}, which represents the user's expert knowledge in the choice domain, and has been shown to influence user reliance and trust in intelligent systems~\cite{hoff2015trust,sanchez2014understanding}.

\begin{figure*}[ht]
    \vspace{-0.3cm}
    \centering
    \includegraphics[width=.8\textwidth]{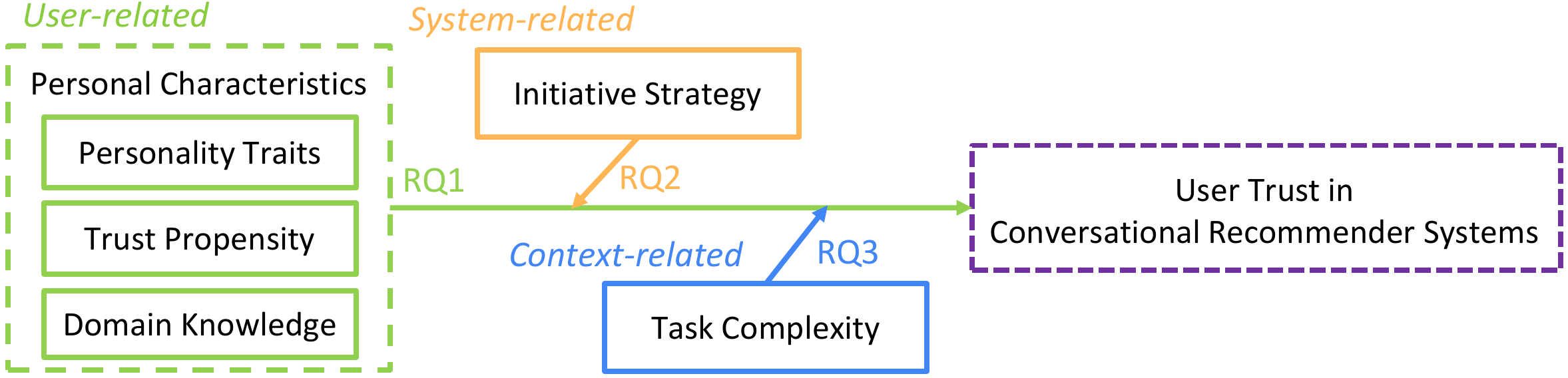}
    \caption{Our research questions.}
    \Description{Figure 1 shows our research questions. The first question is ``How do personal characteristics (personality, trust propensity, domain knowledge) affect user trust in CRSs?'' The second question is ``How do personal characteristics and initiative strategy interact to affect user trust in CRSs?'' The third question is ``How do personal characteristics and task complexity interact to affect user trust in CRSs?''}
    \label{fig:research-question}
    \vspace{-0.3cm}
\end{figure*}

Among system-related factors, we considered \textit{initiative strategies}, among which the \textit{mixed-initiative} strategy is a special characteristic of CRSs. However, it might be difficult to inspire user trust in CRSs with a ``one-size-fits-all'' design of the initiative strategy, because different users may prefer different conversation interactions given their personal characteristics~\cite{Bart:RecSys2011,freitag2016personality}. Thus, we investigated \textit{whether} and \textit{how} users' personal characteristics influence their trust in CRSs with different initiative strategies (\textit{user-initiative} vs. \textit{mixed-initiative}).

Finally, regarding context-related factors (also known as situational factors such as the user's performed task~\cite{knijnenburg2012explaining, hoff2015trust}), we examined \textit{whether} and \textit{how} users' personal characteristics interact with \textit{task complexity} to affect user trust in CRSs. User tasks with different levels of complexity may influence how users interact with systems~\cite{BYSTROM1995191,Myers:CHI2019}, and the influences vary among different types of users~\cite{buckland1991expertise,lai2012influence,Rheu:IJHCI2021}; for example, domain novices tend to spend much more time completing a more complex knowledge-seeking task than do domain experts~\cite{lai2012influence}.

To summarize, we aim to answer the following three research questions in this work (see Figure~\ref{fig:research-question}):

\textbf{RQ1:}  \emph{How do personal characteristics (personality, trust propensity, domain knowledge) affect user trust in CRSs?}

\textbf{RQ2:} \emph{How do personal characteristics and initiative strategy interact\footnote{Here, ``interaction'' is a statistical term. An interaction between A and B to affect Z indicates that A influences Z depending on B, or B influences Z depending on A~\cite{kutner2005applied}.} to affect user trust in CRSs? }

\textbf{RQ3:}  \emph{How do personal characteristics and task complexity interact to affect user trust in CRSs? }

To answer our research questions, we conducted a between-subjects user study (N=148). Two variants of a text-based conversational music recommender were implemented for the experiment: a User-Initiative system that mainly responds to users' requests or feedback, and a Mixed-Initiative system that not only allows users to freely give feedback on the recommendation, but also proactively offers suggestions. Additionally, to vary the task complexity, we designed two user tasks in the context of seeking recommendations: a Simple Task that asks users to find five songs based on their current preferences, and a Complex Task that asks users to first explore diverse types of songs beyond their current interests and then select five songs. 

Our analyses revealed three main findings: 
(1) User experience with conversational interaction in a CRS can be influential on user trust in the system; 
(2) Among the three types of personal characteristics considered, users' \textit{trust propensity} and \textit{domain knowledge} significantly affected user trust toward the CRS;  
(3) The personality trait \textit{conscientiousness} separately interacted with the initiative strategy and the task complexity to inspire user trust in the  CRS. Based on these findings, we present in this paper practical implications for designing trustworthy CRSs that can be tailored to individual users' needs based on their personal characteristics (e.g., \textit{conscientiousness}, \textit{trust propensity}, and \textit{domain knowledge}). We believe this work contributes to the research on conversational Artificial Intelligence (AI) systems, and will facilitate improved CRS design by integrating personalization.

\section{Related Work}

\subsection{Conversational Recommender Systems}

Conversational recommender systems (CRSs) aim to mimic a human advisor to assist users in looking for recommendations in a multi-turn dialogue via text or voice~\cite{jannach2020survey,Christakopoulou:2016,Zhang:CIKM2018,Yang:Recsys2018}, and have been applied in several domains, such as movies~\cite{Cai:UMAP2020}, music~\cite{Jin:CIKM2019,Cai:IUI2021}, and e-commerce~\cite{Zhang:CIKM2018}. Unlike single-shot traditional recommender systems~\cite{Ricci:2010:RSH:1941884}, CRSs allow users to interact with the system in a multi-turn conversation, enabling the system to incrementally refine the user preference model to generate more satisfying recommendations~\cite{Michael:TiiS2017,Cai:UMAP2020}. Such systems can support \textit{mixed-initiative} interaction by combining both \textit{user-initiative} (i.e., users actively tell the system what they want) and \textit{system-initiative} (i.e., the system proactively offers suggestions to users during the recommendation process) interactions, which is regarded as a more flexible interaction strategy in human-computer interaction (HCI)~\cite{Allen:1999}. Several recent studies on CRSs have demonstrated that such systems enable more natural interactions between the user and the system, which can better enhance user experience with recommender systems~\cite{jannach2020survey,Jin:CIKM2019,Cai:IUI2021, narducci2020investigation}.
 
With increased interest in CRSs,  recommender system researchers have been focusing on improving CRS efficiency (i.e., reducing the number of dialogue turns) and effectiveness (i.e., improving the recommendation quality)~\cite{jannach2020survey,GAO2021100, Zhang:CIKM2018}. Although conversational system design is a trending topic within the HCI community, few studies have investigated conversational interaction designs for recommender systems~\cite{Jin:CIKM2019,Cai:IUI2021, Peng:CHI2019, jannach2020survey}. For instance, one study compared two critiquing-based conversational music recommenders that employed different initiative strategies~\cite{Jin:CIKM2019}, and found that the user-initiative CRS gives users more control to tune recommendations on their own, whereas the mixed-initiative CRS guides users to discover more diverse recommendations. Another recent study demonstrated the ability of a CRS to promote user exploration activities~\cite{Cai:IUI2021}, and suggested that the mixed-initiative CRS enhances user exploration by allowing users to control the exploration direction on their own as well as guiding them to explore something different. While existing studies have demonstrated several advantages of CRSs, work on the critical factor of user trust, which strongly determines users' intention to adopt CRSs in real-world situations, is limited.  Thus, this study aims to investigate the factors that may affect user trust in CRSs.

\subsection{Trust in Human-Computer Interaction and Recommender Systems}

Trust is an important factor in both human-human and human-computer relationships~\cite{freitag2016personality, marsh2003role,lee2004trust}, which has been studied for a long period. Trust is defined in various ways in the existing HCI literature~\cite{mcknight1998initial,lee2001trust, wang2005overview}, but a common theme is that trust can be regarded as a behavioral intention (e.g., intention to use) or ``trusting intention''~\cite{mcknight1998initial}. Studies have suggested three types of factors that can influence trust: user-related, system-related, and context-related factors. These three types of factors respectively correspond to the three layers of trust model proposed by Hoff and Bashir~\cite{hoff2015trust}: Dispositional trust refers to the user's general tendency to trust systems, which may arise from individual characteristics such as personality (user-related); learned trust represents the user's evaluations of a system's trustworthiness drawn from past interactions (system-related); situational trust is based on the context of the user-system interaction, such as the complexity of the performed task and user workload (context-related). Motivated by the three-layered trust model~\cite{hoff2015trust}, we are interested in examining these three types of factors (user-related, system-related and context-related) that may influence user trust in CRSs.

Trust-related issues have also gained a lot of attention in recommender systems (RSs), because user trust highly influences users' willingness to use a system and follow its recommendations in their decision-making process~\cite{o2005trust, Komiak:2006,Wang:JMIS2007,Bart:RecSys2011}. User trust in a technological artifact (e.g., recommender system) is often based on competence (i.e., the system's ability to assist users in a specific task), benevolence (i.e., the system's qualities such as security and reputation), and integrity (i.e., the system's reliability and honesty)~\cite{mcknight1998initial}. Studies on RSs have demonstrated that users' perceived competence of the system positively influences their trust in
the system~\cite{Johannes:CHI2019, chen2005trust}. For example, the accuracy and diversity of recommendation lists tend to improve user trust and increase customer purchases in the e-commerce domain~\cite{panniello2016research}. Moreover, the organization-based recommendation interface was demonstrated to reduce user effort in the decision-making process, sustain user trust, and increase users' intention to use the system~\cite{chen2005trust}. Recommendations accompanied by explanations that provide information to assist users in making judgments on the recommended item have also been shown to increase user trust and decision confidence~\cite{tintarev2015explaining, chen2005trust}.

Literature on user trust in RSs  has mostly focused on the aspect of recommendations~\cite{panniello2016research,Johannes:CHI2019,Johannes:CHI2019}, whereas, to the best of our knowledge, user trust in the context of conversational recommendations has rarely been investigated. In CRSs, the conversational interaction between users and the system usually mimics human communication, suggesting that user trust toward the system is similar to trust in interpersonal relationships. Thus, to improve trust, the system should be both reliable in performing the requested tasks and predictable in interactions (i.e., behaving as expected by the user)~\cite{Rheu:IJHCI2021}. However, individual users may have their own expectations of interaction strategies (e.g., preference for \textit{user-initiative} or \textit{mixed-initiative}) depending on their individual characteristics, which may influence their trust in the system. To facilitate the design of trustworthy CRSs that can serve individual users' needs, our work focuses on investigating the impact of personal characteristics on user trust in CRSs that employ different initiative strategies.

\subsection{Personal Characteristics}
\label{sec:personal-characteristics}
Because previous HCI and RS studies have indicated that user trust in the human-system relationship depends on individual characteristics~\cite{cho2016effect, zhou2020effects, hoff2015trust}, we believe that user trust in CRSs may also be influenced by users' personal characteristics. The literature suggests that three personal characteristics, namely \textit{personality traits}, \textit{trust propensity} and \textit{domain knowledge}, are likely to affect user trust in conversational recommenders.


\textbf{Personality Traits.} Personality is defined as individual differences in one's enduring way of thinking, feeling, and behaving~\cite{kazdin2000encyclopedia, McCrae1992Introduction}. The Big-Five personality model, which comprises five traits -- \textit{openness to experience} (openness), \textit{conscientiousness}, \textit{extroversion}, \textit{agreeableness}, and \textit{neuroticism} -- is widely used to assess user personality~\cite{McCrae1992Introduction}. Studies have 
reported the impacts of personality traits on trust in interpersonal relationships~\cite{freitag2016personality}, demonstrating that \textit{openness} and \textit{conscientiousness} affect trust in both friends and strangers, and \textit{agreeableness} affects trust in strangers. Personality traits also influence user trust in the human-machine collaboration~\cite{cho2016effect, zhou2020effects}; for example, people who are more \textit{agreeable} and \textit{conscientious} are more likely to trust automation in decision-making~\cite{cho2016effect}. Thus, we speculate that personality traits (such as \textit{agreeableness} and \textit{conscientiousness}) can also influence user trust toward system guidance in CRSs.

\textbf{Trust Propensity}. \textit{Trust propensity} is defined as the general tendency to trust others~\cite{rotter1971generalized, colquitt2007trust} and is viewed as a dynamic individual difference that may be affected by personality type as well as situational factors (e.g., cultural background)~\cite{mayer1995integrative}. Trust literature has shown that a user's \textit{trust propensity} influences the formation of trust toward specific technological systems~\cite{mcknight1998initial,mcknight2011trust}. When deciding whether to trust a system, users tend to look for cues that signify the system's trustworthiness; however, the perception of the signals is affected by their \textit{trust propensity}~\cite{lee2001trust}. Thus, we seek to determine whether this characteristic will impact user trust in CRSs.

\textbf{Domain Knowledge}. \textit{Domain knowledge} refers to a person's expert knowledge in a specific field. HCI research has demonstrated that users' \textit{domain knowledge} can influence their interaction behaviors and preferred interaction strategies~\cite{nourani2020role}. In recommender systems, domain experts prefer more control during the decision-making process~\cite{Bart:RecSys2011},  whereas domain novices tend to perceive recommendations without too much control to be more helpful. Moreover, users' reliance on decision support systems is related to their \textit{domain knowledge}; for example, users with little or no specialized \textit{domain knowledge} are likely to rely on the system's suggestions~\cite{bussone2015role}. Thus, we believe that \textit{domain knowledge} may influence the way users prefer to interact with CRSs (e.g., preference for \textit{user-initiative} or \textit{mixed-initiative}), hence affecting user trust.

\section{User Experiment}

\begin{figure*}[t]
    \vspace{-0.4cm}
    \centering
    \includegraphics[width=.9\textwidth]{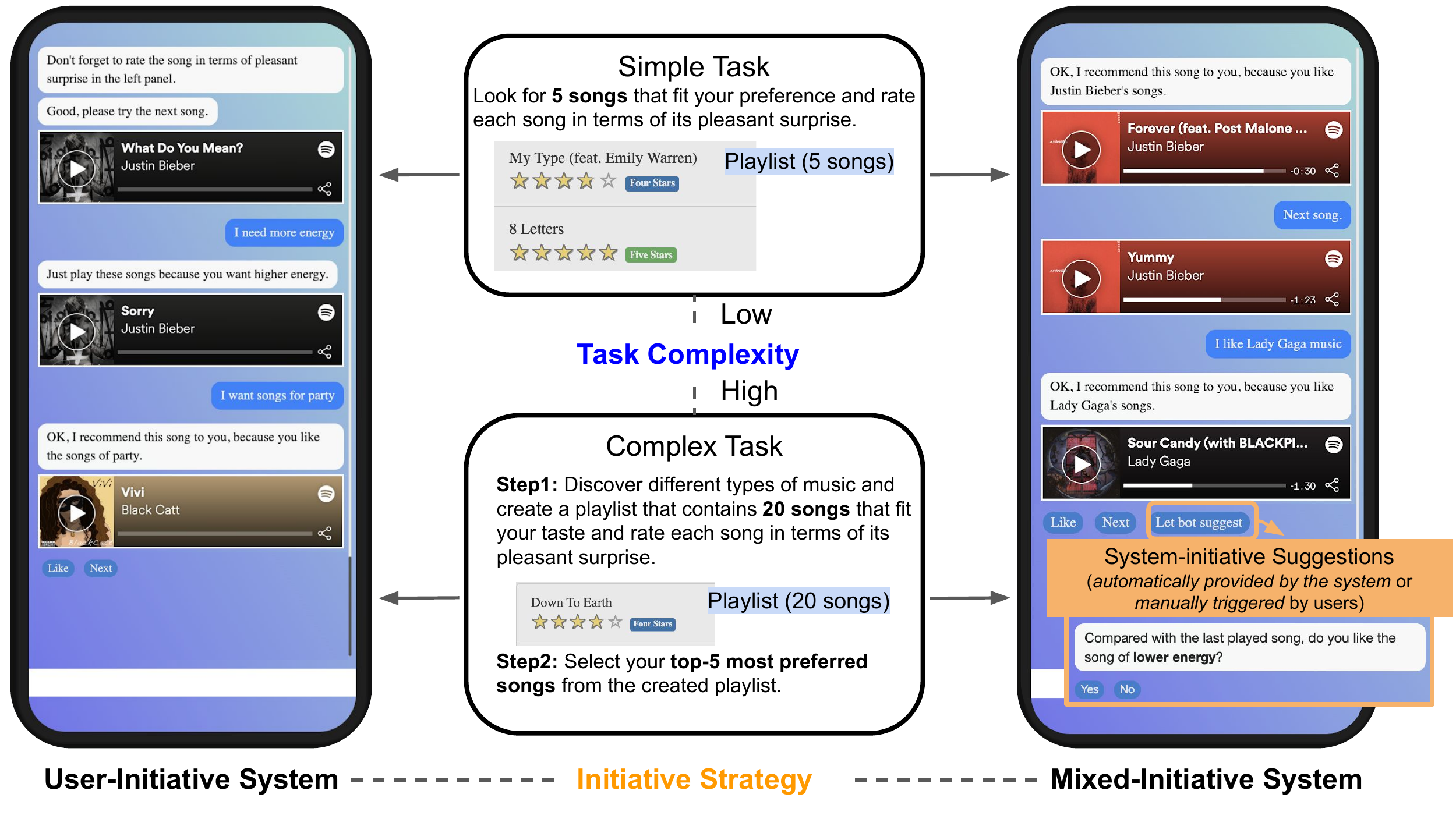}
    \caption{Interfaces of two text-based conversational music recommenders employing different initiative strategies (User-Initiative [left] and Mixed-Initiative [right]), and user tasks with low and high complexity (Simple Task and Complex Task [middle]) in our 2 $\times$ 2 between-subjects study.}
    \Description{Figure 2 shows the interfaces of two text-based conversational music recommenders employing different initiative strategies. The left side of the figure shows the interface of the User-Initiative system, and the right side of the figure shows the interface of the Mixed-Initiative system. The middle shows two user tasks, simple task and complex task in our 2 $\times$ 2 between-subjects study.}
    \label{fig:interface}
    \vspace{-0.4cm}
\end{figure*}

\subsection{Experiment Design}
Based on Hoff and Bashir's three-layered trust model~\cite{hoff2015trust}, we investigated how user-related factors (personal characteristics) interact with both a system-related factor (initiative strategy) and a context-related factor (task complexity) to influence user trust in CRSs. We deployed two text-based prototype conversational music recommenders that employ different initiative strategies (\textit{user-initiative} and \textit{mixed-initiative})~\cite{Cai:IUI2021}, and designed two user tasks of varying complexity in the recommendation domain. Thus, we designed a 2 (User-Initiative vs. Mixed-Initiative) $\times$ 2 (Simple Task vs. Complex Task) online between-subjects user study, in which participants were randomly assigned to one of the four experimental conditions (see Figure~\ref{fig:interface}). Below we present two experimental manipulations.

\subsubsection{Conversational Recommenders}
We used two variants of text-based conversational music recommenders that employ different initiative strategies to support users in looking for music recommendations~\cite{Cai:IUI2021}: 
\begin{itemize}
    \item \textbf{User-Initiative System}: This system, which performs reactive system behavior, only responds when users initiate requests during the conversation. In this system, users can post feedback to refine the current recommended item or ask for songs based on music-related attributes (e.g., genres, tempo, and danceability). For example, a user can tune a recommendation by typing ``\textit{I want higher tempo.}''
    \item \textbf{Mixed-Initiative System}: This system supports both user-initiative and system-initiative interactions. Specifically, in addition to reactively responding to users' requests, the system can proactively provide suggestions (e.g., ``\textit{Compared with the last played song, do you like the song of lower tempo?}'') to facilitate users' music discovery during the recommendation process. As suggested by a study of chatbot proactivity~\cite{Peng:CHI2019}, our system offers suggestions to users when they make an explicit request (i.e., by clicking the ``Let bot suggest'' button; Figure~\ref{fig:interface}) or when the system identifies a good time to offer suggestions.\footnote{According to our pilot test observations, it is reasonable for the system to provide suggestions when the user has consecutively skipped three songs or listened to five songs.} 
\end{itemize}

Although conversational systems can employ three types of initiative strategies, namely user-initiative, system-initiative, and mixed-initiative strategies, we did not employ a purely system-initiative strategy in our study because this design relies on a ``system asks, user responds'' conversation paradigm~\cite{Zhang:CIKM2018}, which can restrict user interaction, reduce flexibility, and make users feel passive~\cite{jannach2020survey,jurafsky2000speech}.

Figure~\ref{fig:interface} shows the user interfaces of the two conversational music recommenders, and the dialogue windows show the conversation between the user and the system. Each recommended song is displayed on a card using which the user can control music playback, along with a set of buttons under the card for the user to give feedback. Specifically, the user can click the ``Like'' button to add the current song into their playlist where they can rate the song, and the ``Next'' button to skip the current song. In the Mixed-Initiative system, the user can click the ``Let bot suggest'' button to trigger the system's suggestion based on the currently recommended song. Additionally, the user can send a message in natural language about the music genre, audio feature, or artist to provide feedback on the currently recommended song and accordingly refine the recommendation. 
We used a popular natural language understanding platform, DialogFlow,\footnote{https://cloud.google.com/dialogflow/es/docs} and a widely used online music service, Spotify API,\footnote{https://developer.spotify.com/documentation/web-api} to develop our conversational music recommenders. For the generation of the system-initiative suggestions, we employed the progressive system-suggested critiquing technique designed by Cai et al.~\cite{Cai:IUI2021}, which considers the user's song preferences as well as incremental feedback captured from past interactions.

\subsubsection{User Tasks}
\label{sec:user-task}
To determine whether and how users' personal characteristics interact with the context-related factor (task complexity) to influence user trust in CRSs, we considered two typical user tasks in the recommendation domain:

\begin{itemize}
    \item \textbf{Simple Task}. Users are asked to interact with our conversational music recommender (called ``music chatbot'' in our study)  to find five songs that suit their preferences, and rate each song in terms of its \textit{pleasant surprise}. 
    \item \textbf{Complex Task}. Users are asked to complete two steps: (1) use our music chatbot to discover songs as many different music genres as possible, create a playlist containing 20 songs that fit their tastes, and then rate each song in terms of its \textit{pleasant surprise}; and (2) select their top-5 most preferred songs from the playlist they created. Compared with the simple task, this task requires users to discover more types of music and make comparisons for selecting their most preferred songs, which is more cognitively demanding. 
\end{itemize}

\subsection{Participants}
We recruited participants from Prolific,\footnote{https://www.prolific.co/} a popular platform for academic surveys~\cite{peer2017beyond}. To ensure experiment quality, we pre-screened users in Prolific using the following criteria: (1) participants should be fluent in English; (2) they must have more than 100 previous submissions; (3) their approval rate should be greater than 95\%. The experiment took 25 minutes to complete on average. We compensated each participant \pounds 2.40  on successfully completing the experiment. The Research Ethics Committee (REC) of the authors' university approved this study. 

In total, 194 users participated in our study. We removed the responses of 23 participants because of their excessively long experiment completion time (outliers). We excluded the responses of another 23 participants who failed the attention check questions.\footnote{To ensure the quality of user responses, we set three attention checking questions (e.g.,``\textit{Please indicate which of the following items is not fruit?'').}} Thus, the remaining responses of 148 participants were included in the analyses [User-Initiative: Simple Task (32), Complex Task (35); Mixed-Initiative: Simple Task (45), Complex Task (36); Gender: female (70), male (75), other (3); Age: 19-25 (69), 26-30 (27), 31-35 (25), 36-40 (10), 41-50 (11), > 50 (6)]. Participants were mainly from the United Kingdom (32), the United States (32), Portugal (18),  Poland (12), and Italy (9).


\subsection{Experimental Procedure}
Participants had to accept a general data protection regulation consent form before they signed into our system using their Spotify accounts. After reading the user study instructions, participants were asked to fill out a pre-study questionnaire, which included demographic questions and questions for measuring their personal characteristics (see Section~\ref{sec:pre-study-questionnaire}). To ensure that participants understood the study task and how to use the conversational recommender, they were given a tutorial of interacting with the assigned conversational music recommender, followed by two minutes to try the system. After completing the tutorial, participants were asked to complete a randomly assigned task (Simple Task or Complex Task as described in Section~\ref{sec:user-task}). After finishing the task, participants were asked to fill out a post-study questionnaire regarding their trust-related perception of the conversational music recommender (see Section~\ref{sec:measure-questionnaires}).

\subsection{Pre-Study Questionnaire}
\label{sec:pre-study-questionnaire}
In the pre-study questionnaire, we used a short personality test, the Ten Item Personality Inventory (TIPI)~\cite{gosling2003very}, to assess participants' Big-Five \textit{personality traits}: \textit{openness to experience}, \textit{conscientiousness}, \textit{extroversion}, \textit{agreeableness}, and \textit{neuroticism}.  Each personality trait is assessed by two questions in the TIPI, and the personality value for each trait is the average of the scores on the two questions. To measure participants' \textit{trust propensity}, we adopted two statements developed by Lee and Turban~\cite{lee2001trust}: \textit{``I tend to trust the recommender, even though having little knowledge of it.''} and \textit{``Trusting someone or something is difficult.''} Because our system was built for the music domain, we used the nine statements from the ``Active Musical Engagement'' facet of Goldsmiths Musical Sophistication Index~\cite{mullensiefen2014musicality} to assess participants' \textit{musical sophistication} as their \textit{domain knowledge}. All statements were rated on a 7-point Likert scale from 1 (\textit{strongly disagree}) to 7 (\textit{strongly agree}). In Table~\ref{tab:pc_explanation}, we briefly introduce each measured personal characteristic.

Table~\ref{tab:user-characteristics-statistics} shows the descriptive statistics of our participants' personal characteristics (PCs). The scored values are centered between 3 and 5 for almost all PCs, and the standard deviations are comparable across all PCs. Table~\ref{tab:user-characteristics-correlation} shows Pearson's correlations between these PCs; these correlations (e.g., \textit{trust propensity} is positively related to \textit{extroversion} and \textit{agreeableness}) are generally consistent with the results of previous literature~\cite{greenberg2015personality, john1999big, freitag2016personality}.


\begin{table*}[t]
\vspace{-0.2cm}
\footnotesize
\caption{Description of Big-Five personality traits, trust propensity, and domain knowledge (musical sophistication)}
\label{tab:pc_explanation}
\begin{tabular}{p{0.2cm}lp{24pc}}
\toprule
\multicolumn{2}{l}{\textbf{Personal Characteristic (PC)}} & \textbf{Description} 
\\ \midrule
\multicolumn{2}{l}{\textbf{Big-Five Personality Traits~\cite{gosling2003very,greenberg2015personality}}} &  \\   \midrule
& \textbf{Openness to Experience (O)} & This trait, also called \textbf{Openness}, is related to one's cognitive style, distinguishing \newline creative, imaginative people (high O) from down-to-earth, conventional people (low O). \\   \cmidrule(l){2-3} 
&  \textbf{Conscientiousness (C)} & This trait is associated with one's way of controlling, regulating, and directing impulses, \newline  distinguishing prudent people (high C) from impulsive people (low C). \\  \cmidrule(l){2-3} 
&  \textbf{Extroversion (E)} &  This trait concerns the active level of engagement with the external world, distinguishing \newline sociable, outgoing people (high E) from reserved, quiet people (low E). \\  \cmidrule(l){2-3} 
&  \textbf{Agreeableness (A)} & This trait reflects one's attitude toward cooperation and social harmony, distinguishing \newline cooperative, sympathetic people (high A) from critical, tough people (low A). \\  \cmidrule(l){2-3} 
&  \textbf{Neuroticism (N)} &   This trait describes one's tendency to experience negative feelings,  distinguishing \newline sensitive, easily upset people (high N) from calm, unflappable people (low N). \\ \midrule
\multicolumn{2}{l}{\textbf{Trust Propensity (TP)~\cite{lee2001trust}}} &  TP reflects one's general willingness to trust other people or technologies. \newline People with high TP are naturally inclined to trust others, while people with low TP are hesitant. \\ \midrule
\multicolumn{2}{l}{\textbf{Musical Sophistication (MS)~\cite{mullensiefen2014musicality}}}&  MS is related to one's ability to successfully engage with music. People with high MS are \newline more flexible in responding to a great range of musical situations than are people with low MS.    \\
\bottomrule
\end{tabular}
\vspace{-0.2cm}
\end{table*}

\begin{table}[t]
    \footnotesize
    \centering
    \captionof{table}{Descriptive statistics of participants' personal characteristics (PCs)}
    \label{tab:user-characteristics-statistics}
    \begin{tabular}{lccccc}
    \toprule
    \textbf{PC} & \textbf{Min} & \textbf{Median} & \textbf{Mean} & \textbf{Max} & \textbf{S.D.} \\ \midrule
    \textbf{O} & 2.00 & 5.00 & 5.01 & 7.00 & 1.15 \\
    \textbf{C} & 2.00 & 5.25 & 5.19 & 7.00 & 1.19 \\
    \textbf{E} & 1.00 & 3.25 & 3.29 & 7.00 & 1.54 \\
    \textbf{A} & 2.00 & 5.00 & 4.94 & 7.00 & 1.10 \\
    \textbf{N} & 1.00 & 3.50 & 3.51 & 6.50 & 1.53 \\
    \textbf{TP} & 1.00 & 4.00 & 4.05 & 6.50 & 0.99 \\
    \textbf{MS} & 1.44 & 4.22 & 4.25 & 6.89 & 1.03 \\
     \bottomrule
    \end{tabular}
\vspace{-0.3cm}
\end{table}
\begin{table}[t]
    \footnotesize
    \centering
    \captionof{table}{Pearson's correlations between the Big-Five personality traits, trust propensity, and musical sophistication}
    \label{tab:user-characteristics-correlation}
    \begin{tabular}{llllllll}
    \toprule
    \textbf{PC} & \textbf{O} & \textbf{C} & \textbf{E} & \textbf{A} & \textbf{N} & \textbf{TP} & \textbf{MS} \\ \midrule
    \textbf{O} & - & *** & ** & *** & *** &  & *** \\
    \textbf{C} & 0.2858 & - &  & *** & *** & * &  \\
    \textbf{E} & 0.2189 & 0.0894 & - &  & ** & ** & * \\
    \textbf{A} & 0.2920 & 0.3321 & 0.1518 & - & *** & *** &  \\
    \textbf{N} & -0.3112 & -0.3277 & -0.2375 & -0.3954 & - &  &  \\
    \textbf{TP} & 0.1419 & 0.1854 & 0.2668 & 0.2729 & -0.1509 & - &  \\
    \textbf{MS} & 0.2875 & 0.0702 & 0.2086 & 0.0213 & 0.0326 & 0.0916 & - \\
    \bottomrule
    \multicolumn{8}{l}{Significance: *** $p$ < .001, ** $p$ < .01, * $p$ < .05.} 
    \end{tabular}
\vspace{-0.3cm}
\end{table}

\subsection{Trust Measurement}
\label{sec:measure-questionnaires}

In the post-study questionnaire, we measured users' trust-related perception of the conversational music recommender in two main dimensions: Competence Perception and User Trust. Competence Perception refers to how users perceive the system's competence in assisting them in performing tasks, which contains the following three constructs derived from prior works~\cite{Chen:aaai2006,knijnenburg2012explaining,walker1997paradise}: 
\begin{itemize}
    \item \textit{Perceived Recommendation Quality}: This construct measures the system's ability to provide good recommendations to help users make decisions or support their exploration. Users may judge the quality of recommendations in terms of several aspects, e.g., accuracy, novelty, and serendipity~\cite{Chen:aaai2006,knijnenburg2012explaining}. A previous study showed that users' perceived recommendation quality influences their perceived usefulness of the system in helping them accomplish tasks, which consequently impacts user trust toward the system~\cite{Chen:aaai2006}. Thus, we considered this construct and measured it using questions from ResQue~\cite{Chen:aaai2006}, a widely used user-centric evaluation framework for recommender systems.
    
    \item \textit{Perceived Conversational Interaction}: This construct measures the system's ability to effectively communicate with users to perform tasks during the interaction. Several aspects of conversational interaction are deemed crucial to CRSs~\cite{jin2021key}, which include understandability, perceived control, interaction adequacy (i.e., ability to elicit and refine preferences~\cite{Chen:aaai2006}), and naturalness of the dialogue interaction. Because communication is the primary way people develop trust within interpersonal relationships~\cite{de2013communication}, we hypothesize that users' experience with conversational interaction will also influence the formation of user trust in the system. We measured this construct by adopting questions mainly from an evaluation framework for conversational agents~\cite{walker1997paradise}.
    
    \item \textit{Perceived Effort}: This construct measures users' perceived difficulty or ease in using the system for completing their tasks, which can reflect the effectiveness of the system in supporting users to accomplish tasks. When users perceive high effort in using the system to complete tasks, they may feel frustrated and show less trust~\cite{Chen:aaai2006,chen2005trust}. We used questions in ResQue~\cite{Chen:aaai2006} to measure this construct.  
\end{itemize}

The User Trust dimension directly measures user trust in the CRS based on two constructs, each measured using one question item: \textit{Perceived Trust} assesses users' overall feelings of trust toward the conversational recommender, and \textit{Intention to Use} measures users' willingness to use the system in the future.

We assessed the validity of our constructs as measured by the question items (19 items in the initial questionnaire) by conducting confirmatory factor analysis (CFA) with R library Lavvan.\footnote{http://lavaan.ugent.be/} In CFA, the items within the same scale are represented by a latent factor, where the loading of each item denotes how strongly that item is associated with the corresponding factor. We iteratively removed 5 items with low loadings (<0.50) or high cross-loadings, leaving behind 14 items in total (Table~\ref{tab:questionnaire}). All items were assessed by 7-point Likert scale from 1 (\textit{strongly disagree}) to 7 (\textit{strongly agree}). Each factor had good internal consistency (Cronbach's $\alpha$ > 0.80), composite reliability (CR > 0.80), and convergent validity [Average Variance Extracted (AVE) > 0.50]~\cite{ab2017discriminant}, and the loading of each item exceeded the acceptable level of 0.50, with an overall good model fit~\cite{hu1999cutoff}: $\chi^2$(51) = 86.283, $p$ < .001; Root Mean Square Error of Approximation (RMSEA) = 0.068,   Comparative Fit Index (CFI) = 0.967,  Turker-Lewis
Index (TLI) = 0.957.

\begin{table*}[t]
\vspace{-0.3cm}
\footnotesize
\caption{Post-study questionnaire for measuring users' trust-related perception of the conversational recommender}
\label{tab:questionnaire}
\begin{tabular}{lp{24pc}l}
\toprule
\textbf{Construct} & \textbf{ Item (each statement is rated on a 7-point Likert scale)} & \textbf{Loadings} \\ \midrule
\multicolumn{3}{l}{\textbf{Competence Perception} }  \\  \midrule
\multicolumn{3}{l}{\textit{\textbf{Perceived Recommendation Quality}} (Cronbach alpha: 0.9001; CR: 0.8951; AVE: 0.6647)} \\
 & The music chatbot helped me discover new songs. & 0.7940 \\
 & The songs recommended to me were novel. & 0.5378 \\
 & The music chatbot provided me with recommendations that I had not considered in the first place \newline\hspace*{10pt} but turned out to be a positive   and surprising discovery. & 0.8457 \\
  & The music chatbot provided me with surprising recommendations that helped me discover new songs \newline\hspace*{10pt} that I wouldn't have found elsewhere. & 0.9226 \\
   & The music chatbot provided me with recommendations that were a pleasant surprise to me \newline\hspace*{10pt} because I would not have discovered them somewhere else.  & 0.8728 \\
\multicolumn{3}{l}{\textit{\textbf{Perceived Conversational Interaction}} (Cronbach alpha: 0.8668; CR: 0.8692; AVE: 0.5756)} \\
 & I   found the music chatbot easy to understand in this conversation. & 0.7590 \\
 & The   music chatbot worked the way I expected it to in this conversation. & 0.7950 \\
 & I   found it easy to inform the music chatbot if I dislike/like the recommended   song. & 0.6967 \\
 & I   felt in control of modifying my taste using this music chatbot. & 0.7995 \\
  & In this conversation, I knew what I could say or do at each point of the dialog.& 0.7236 \\
\multicolumn{3}{l}{\textit{\textbf{Perceived Effort}} (Cronbach alpha: 0.8712; CR: 0.8730; AVE: 0.7729)} \\
 & Looking   for a song using this interface required too much effort.  & 0.8675 \\ 
 & I   easily found the songs I was looking for. (reversed) & 0.8927 \\ \midrule
\multicolumn{3}{l}{\textbf{User Trust}} \\ \midrule
\textit{\textbf{Perceived Trust}} & This music chatbot can be trusted. &  \\
\textit{\textbf{Intention to Use}} & I   will use this music chatbot again. &  \\ \bottomrule
\end{tabular}
\end{table*}

\section{Analyses \& Results} 
The three-layered trust model~\cite{hoff2015trust} indicates three types of factors that may influence user trust: user-related, system-related, and context-related factors. We conducted a series of analyses to investigate the influences of these factors on users' trust-related perception of CRSs. First, we examined the relationship between Competence Perception and User Trust, and the impacts of user-related factors (i.e., the three personal characteristics) on these two dimensions (RQ1). For this purpose, we used structural equation modeling (SEM) to build a path model to test and evaluate multivariate causal relationships among the constructs in Table~\ref{tab:questionnaire} and the effects of personal characteristics in an integrative structure. 

Next, we investigated in-depth the impacts of personal characteristics to determine whether and how user-related factors interact with the system-related factor (initiative strategy) and the context-related factor (task complexity) to influence Competence Perception and User Trust (RQ2 \& RQ3). As it is relatively complicated to perform interaction effect analyses with multiple factors using SEM~\cite{henseler2010comparison}, we conducted an additional set of linear regression analyses to investigate the interaction effects.

\begin{figure*}[!ht]
    \vspace{-0.2cm}
    \centering
    \includegraphics[width=.88\textwidth]{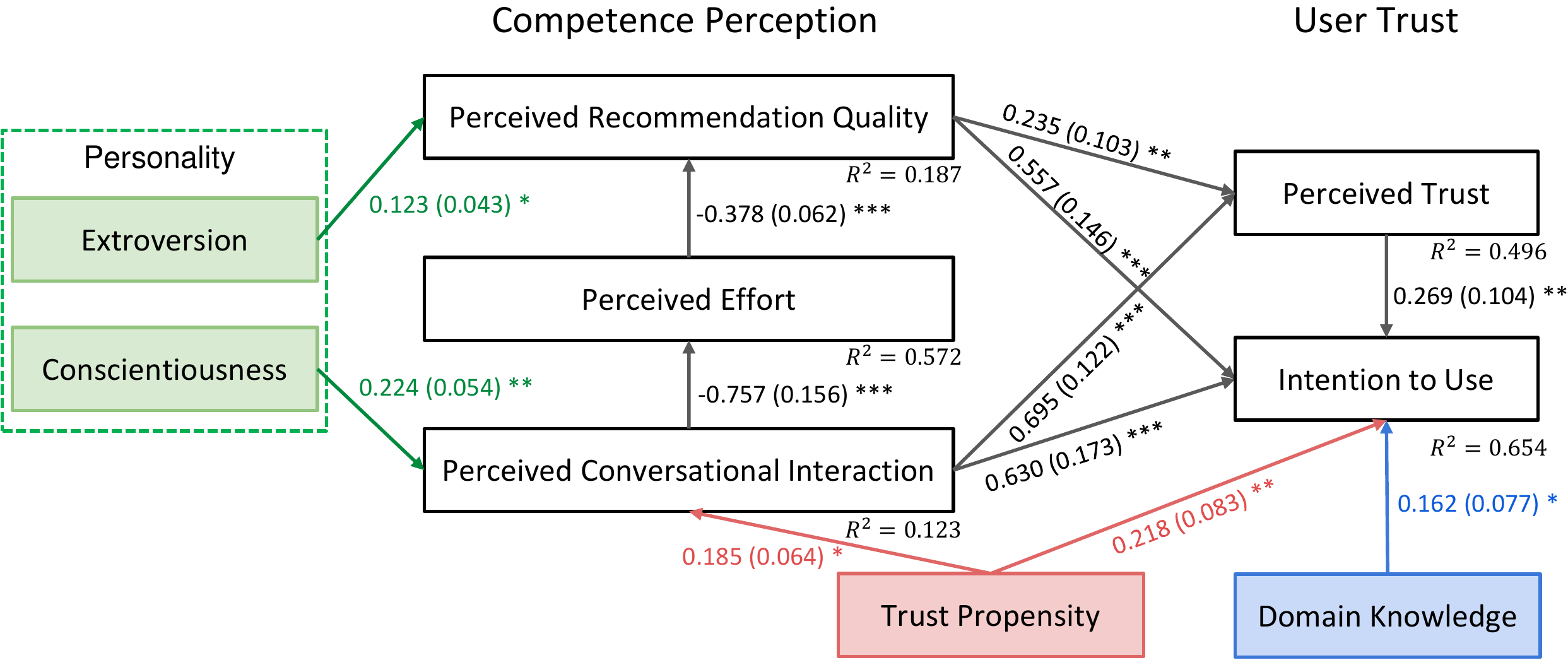}
    \vspace{-0.2cm}
    \caption{Structural equation modeling (SEM) results. Two personality traits (\textit{conscientiousness} and \textit{extroversion}) influenced User Trust via Competence Perception, and \textit{trust propensity} and \textit{musical sophistication} directly affected User Trust. The numbers on the arrows represent the $\beta$ coefficient and standard error (in parentheses) of the effect. Significance: *** $p$ < .001, ** $p$ < .01, * $p$ < .05. $R^2$ is the proportion of variance explained by the model. Factors are scaled to have a standard deviation of 1.}
    \Description{Figure 3 shows the results of structural equation modeling. The results indicate that two personality traits, conscientiousness and extroversion, influenced User Trust via Competence Perception, and trust propensity and musical sophistication directly affected User Trust. The numbers on the arrows represent the $\beta$ coefficient and standard error (in parentheses) of the effect.}
    \label{fig:SEM-Trust}
    \vspace{-0.2cm}
\end{figure*}

\begin{table*}[t]
\vspace{-0.3cm}
\footnotesize
\caption{Regression models for estimating the interaction effects of personal characteristics with initiative strategy and task complexity on users' trust-related perception constructs (as shown in Table~\ref{tab:questionnaire}) in the conversational recommender}
\label{tab:regression}
\begin{tabular}{lccccc}
\toprule
 &  \textbf{Perceived} & \textbf{Perceived} & & & \\ 
 &  \textbf{Recommendation  Quality} & \textbf{Conversational  Interaction} &\textbf{Perceived Effort} & \textbf{Perceived Trust} & \textbf{Intention to Use} \\ 
 &  Coef. (S.E.)&  Coef. (S.E.) &  Coef. (S.E.) &  Coef. (S.E.) &  Coef. (S.E.) \\ \midrule
Mixed Initiative vs. User Initiative & 0.408 (0.229) . & -0.064 (0.147) & -0.183 (0.235) & 0.054 (0.190) & -0.003 (0.259) \\
Complex Task vs. Simple Task & 0.222 (0.228) & -0.290 (0.146) * & 0.391 (0.234) . & -0.055 (0.189) & -0.408 (0.258) \\
Openness & 0.086 (0.199) & -0.053 (0.128) & -0.073 (0.205) & 0.076 (0.165) & -0.143 (0.225) \\
Conscientiousness & -0.038 (0.182) & 0.133 (0.117) & 0.051 (0.187) & -0.055 (0.151) & 0.054 (0.207) \\
Extroversion & -0.099 (0.149) & -0.092 (0.095) & 0.050 (0.153) & 0.165 (0.123) & -0.086 (0.168) \\
Agreeableness & 0.131 (0.199) & -0.151 (0.128) & 0.078 (0.205) & -0.088 (0.165) & 0.162 (0.226) \\
Neuroticism & 0.141 (0.153) & -0.069 (0.098) & -0.220 (0.157) & 0.136 (0.127) & -0.004 (0.173) \\
Trust Propensity & 0.189 (0.248) & 0.076 (0.159) & -0.122 (0.255) & 0.029 (0.206) & 0.212 (0.281) \\
Musical Sophistication & 0.782 (0.221) *** & 0.189 (0.142) & -0.399 (0.227) . & 0.280 (0.183) & 0.620 (0.250) * \\
Mixed Initiative x Openness & -0.013 (0.230) & 0.241 (0.148) & -0.218 (0.237) & 0.066 (0.191) & 0.128 (0.261) \\
Mixed Initiative x Conscientiousness & \textbf{0.652 (0.208) **} & \textbf{0.284 (0.134) *} & -0.388 (0.214) . & \textbf{0.372 (0.173) *} & 0.405 (0.236) . \\
Mixed Initiative x Extroversion & 0.067 (0.173) & -0.064 (0.111) & -0.061 (0.178) & -0.239 (0.143) . & 0.130 (0.196) \\
Mixed Initiative x Agreeableness & -0.327 (0.239) & 0.167 (0.154) & -0.196 (0.246) & -0.016 (0.198) & 0.047 (0.271) \\
Mixed Initiative x Neuroticism & -0.049 (0.175) & 0.175 (0.113) & 0.002 (0.180) & -0.087 (0.145) & 0.179 (0.199) \\
Mixed Initiative x Trust Propensity & -0.224 (0.261) & -0.074 (0.167) & 0.199 (0.268) & -0.040 (0.216) & -0.034 (0.296) \\
Mixed Initiative x Musical Sophistication & \textbf{-0.496 (0.236) *} & -0.143 (0.151) & 0.377 (0.243) & 0.069 (0.196) & 0.010 (0.267) \\
Complex Task x Openness & -0.169 (0.221) & -0.104 (0.142) & 0.324 (0.227) & -0.055 (0.183) & -0.002 (0.250) \\
Complex Task x Conscientiousness & \textbf{-0.526 (0.211) *} & -0.210 (0.136) & 0.184 (0.217) & 0.005 (0.175) & -0.364 (0.239) \\
Complex Task x Extroversion & 0.205 (0.165) & 0.041 (0.106) & 0.050 (0.170) & -0.076 (0.137) & 0.047 (0.187) \\
Complex Task x Agreeableness & 0.354 (0.240) & 0.151 (0.154) & -0.206 (0.247) & 0.304 (0.199) & 0.009 (0.272) \\
Complex Task x Neuroticism & -0.048 (0.167) & -0.008 (0.107) & 0.140 (0.172) & 0.042 (0.138) & -0.020 (0.189) \\
Complex Task x Trust Propensity & 0.337 (0.260) & 0.326 (0.167) . & \textbf{-0.545 (0.267) *} & 0.220 (0.215) & \textbf{0.596 (0.294) *} \\
Complex Task x Musical Sophistication & \textbf{-0.524 (0.242) *} & 0.052 (0.155) & -0.064 (0.249) & -0.190 (0.200) & -0.417 (0.274) \\
Constant & 3.948 (0.207) *** & 5.920 (0.133) *** & 2.840 (0.213) *** & 5.421 (0.172) *** & 5.054 (0.235) *** \\ \midrule
$R^2$ & 0.314 & 0.314 & 0.243 & 0.249 & 0.301 \\
Adjusted $R^2$ & 0.186 & 0.187 & 0.102 & 0.110 & 0.171\\ \bottomrule
\multicolumn{6}{p{41pc}}{Given that interaction effects are present in our regression models, we only interpret the interaction effects (highlighted in bold) because the  interpretation of the main effects\newline  (i.e.,  the effect of one independent variable on the dependent variable) is incomplete or misleading~\cite{kutner2005applied}.\newline Significance: *** $p$ < .001, ** $p$ < .01, * $p$ < .05, . $p$ < .1; Coef. stands for coefficient; S.E. stands for standard error.}
\end{tabular}
\vspace{-0.2cm}
\end{table*}

\subsection{User Trust in Conversational Recommender Systems}
\label{sec:SEM}

Figure~\ref{fig:SEM-Trust} illustrates the results of the structural equation modeling (SEM) analysis, showing all significant paths in our model. The SEM model had overall good model fit indices: $\chi^2$(123) = 182.312, p < .001; RMSEA = 0.057,  CFI = 0.956, TLI = 0.947, which meet the recommended SEM fit standard.\footnote{Hu and Bentler~\cite{hu1999cutoff} suggest good values for the following indices: CFI > .96, TLI > .95, and RMSEA < .05.}

In the resulting model, the paths between the perception constructs (inside black rectangles) show how users' perceptions of the system's competence influenced their trust in the CRS. Specifically, the significant paths (Perceived Recommendation Quality $\rightarrow$ Perceived Trust and Intention to Use; Perceived Conversational Interaction $\rightarrow$ Perceived Trust and Intention to Use) justify the positive effects of users' competence perception of the CRS on their trust in the CRS. Furthermore, the path coefficients indicate that Perceived Trust was affected more by Perceived Conversational Interaction (coefficient = 0.695) than Perceived Recommendation Quality (coefficient = 0.235). Our model also verifies the positive effect of Perceived Trust on Intention to Use~\cite{pu2011user}.  Additionally, we observed an interesting path (Perceived Conversational Interaction $\rightarrow$ Perceived Effort $\rightarrow$ Perceived Recommendation Quality), showing that users' perceptions of conversational interaction positively influenced their perceptions of the recommendation quality, which were mediated by their perceived effort. These effects highlight the importance of considering Perceived Conversational Interaction for inspiring user trust in CRSs.

Moreover, our SEM model shows how personal characteristics influence the constructs of Competence Perception and User Trust. The results indicate that two personality traits (\textit{conscientiousness} and \textit{extroversion}) influenced User Trust via Competence Perception, whereas \textit{trust propensity} and \textit{domain knowledge (musical sophistication)} directly affected User Trust in the CRS. 
\begin{itemize}
    \item \textbf{Conscientiousness}. The trait \textit{conscientiousness} positively influenced users' perceptions of conversational interaction: users with higher \textit{conscientiousness} tended to have a better perception of their interaction with the conversational recommender.

    \item \textbf{Extroversion}. The trait \textit{extroversion} was positively related to users' perceived recommendation quality. Users with higher \textit{extroversion} tended to perceive higher system competence in recommending satisfying songs. One possible explanation is that compared with introverted users, extroverted users (who are more outgoing and vigorous~\cite{john1999big}) are more willing to take risks and try listening to different music during the interaction, hence improving their perceptions of recommendations.
    
    \item \textbf{Trust Propensity}.  \textit{Trust propensity} positively affected users' perceptions of the conversational interaction and their intention to use. Namely, users who are more willing to trust others tended to enjoy the conversational interaction with the CRS and have a higher intention to use it again. People with a higher \textit{trust propensity} (who tend to believe others are sincere and have good intentions~\cite{colquitt2007trust}) may be more cooperative~\cite{jacquet2019impact} with the system during the conversation, resulting in a more positive conversational experience.
    \item \textbf{Musical Sophistication}. Regarding the influence of \textit{domain knowledge}, we found that \textit{musical sophistication} positively influenced users' intention to use the CRS, suggesting that users with higher \textit{musical sophistication} are more likely to use the conversational recommender in the future.  
\end{itemize}

In addition to the user-related factors (personal characteristics), we investigated whether the system-related factor (initiative strategy) and the context-related factor (task complexity) directly influenced user trust in the model. Among these factors, task complexity negatively affected users' perceived conversation interaction ($p$ < .05), which may be attributed to the increased user effort required to perform a complex task.

\subsection{Interaction Effects on User Trust}

\begin{figure*}[t]
    \vspace{-0.4cm}
    \centering
    \subfigure[Interaction effect of   \textit{conscientiousness} and   \textit{initiative}  \textit{strategy}  on  Perceived Recommendation Quality.]
    {\includegraphics[width=0.35\textwidth]{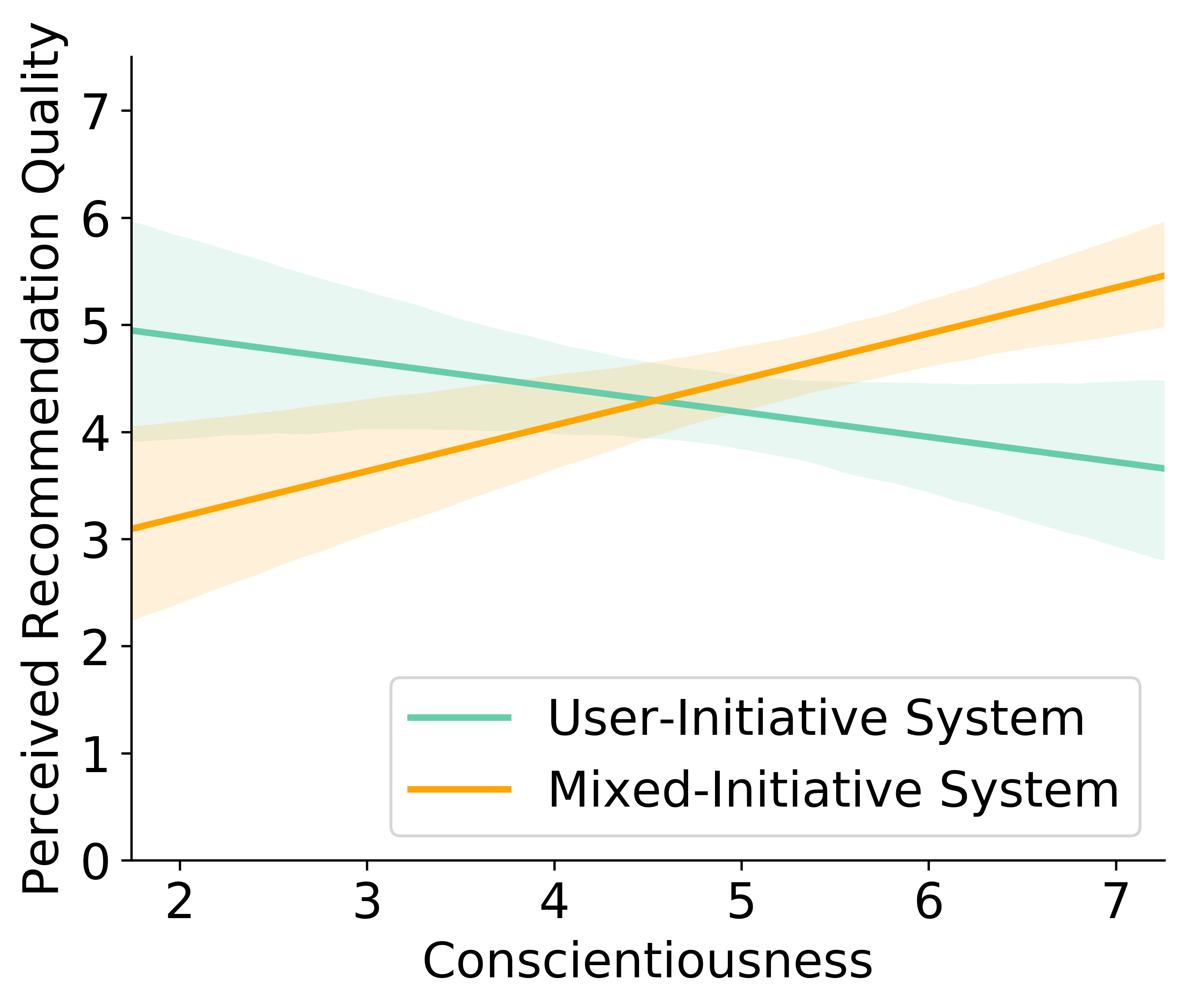} 
    \label{fig:interaction-system-1}}
    \hspace{2cm}
    \subfigure[Interaction effect of   \textit{conscientiousness} and  \textit{initiative}  \textit{strategy}  on  Perceived Conversational Interaction.]
    {\includegraphics[width=0.35\textwidth]{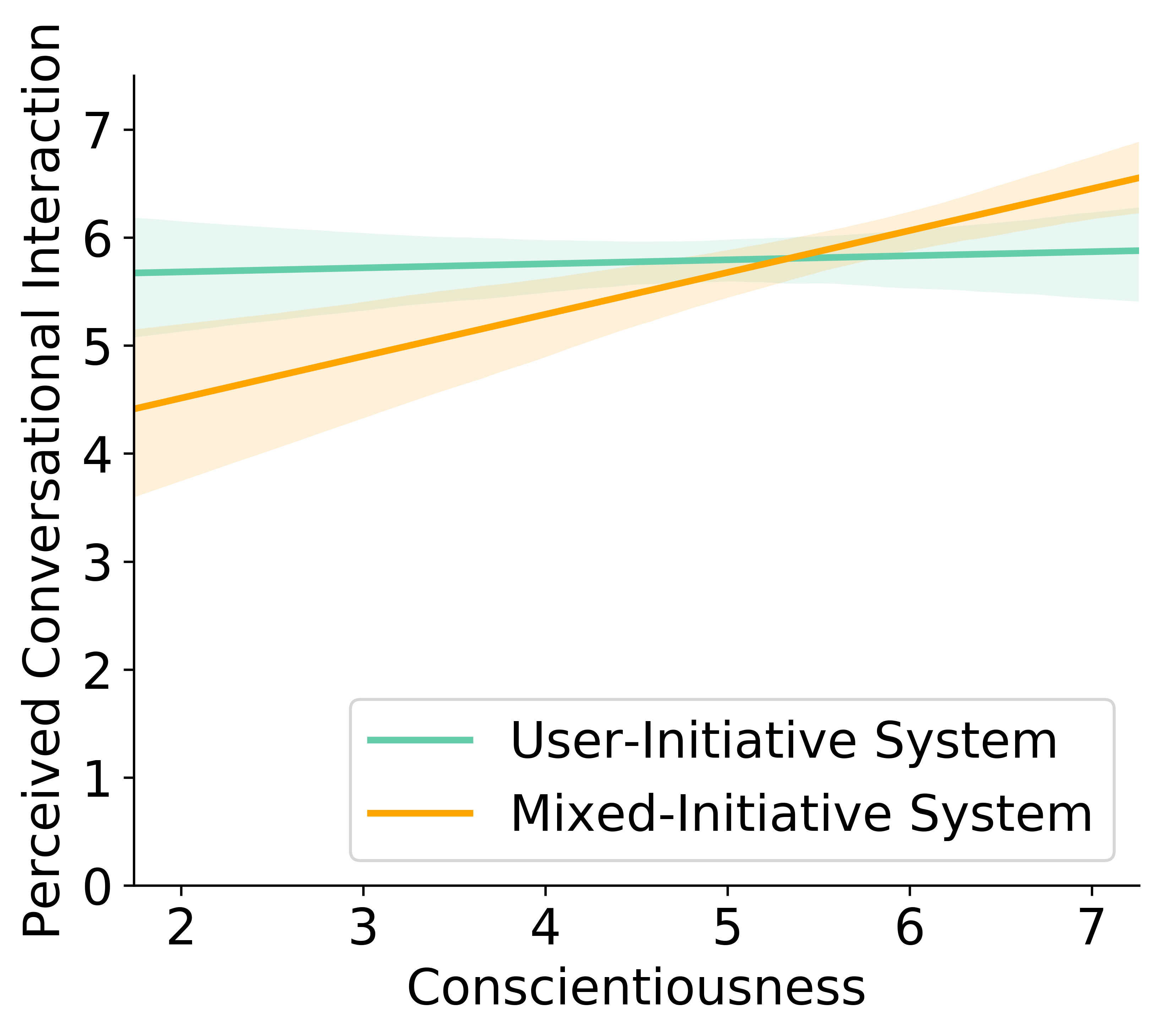} 
    \label{fig:interaction-system-2}}\\
    \vspace{-0.3cm}
    \subfigure[Interaction effect of  \textit{conscientiousness} and  \textit{initiative} \textit{strategy} on Perceived Trust.]
    {\includegraphics[width=0.35\textwidth]{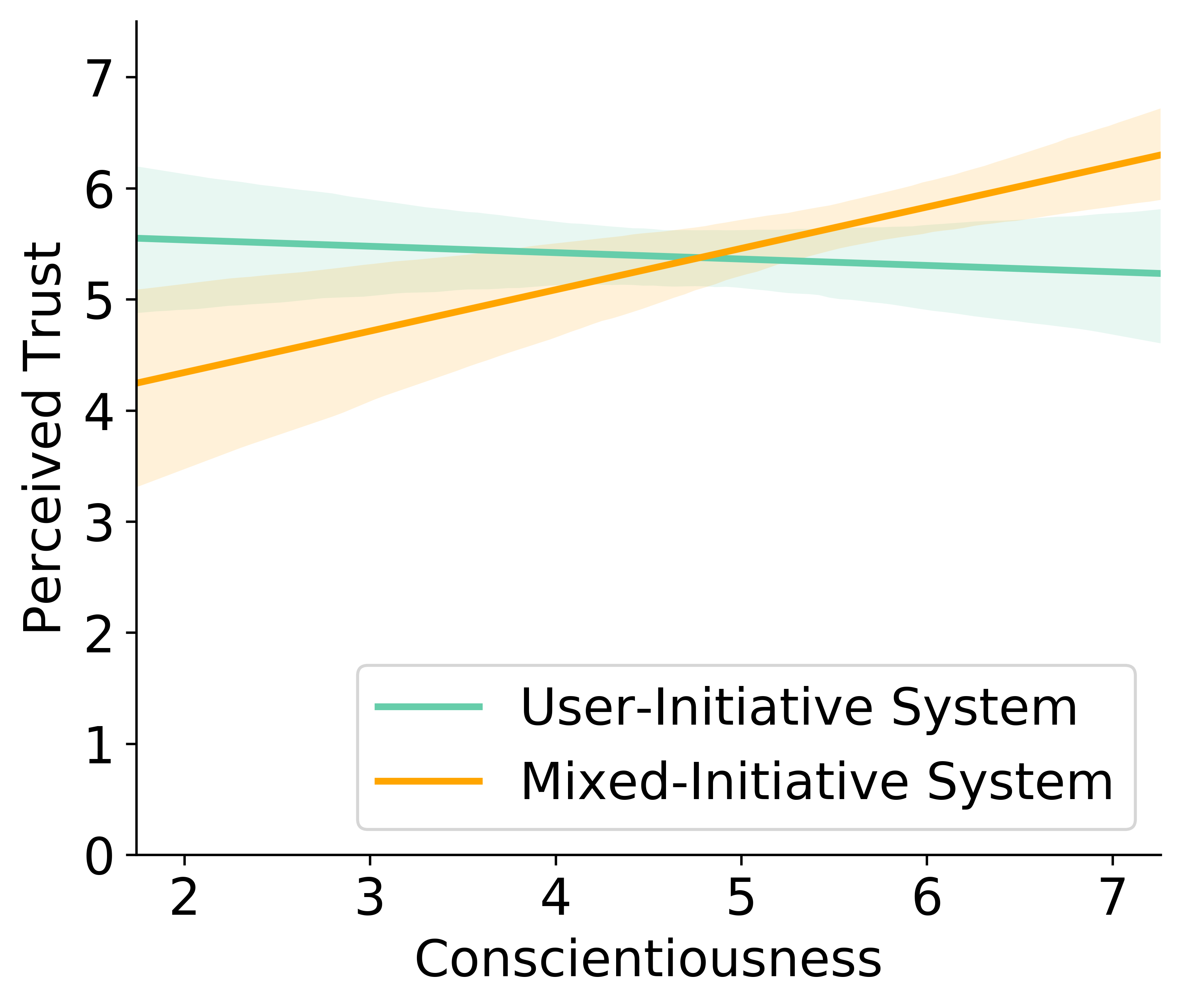} 
    \label{fig:interaction-system-3}}
     \hspace{2cm}
    \subfigure[Interaction effect of   \textit{musical}  \textit{sophistication} and  \textit{initiative}  \textit{strategy} on   Perceived    Recommendation Quality.]
    {\includegraphics[width=0.35\textwidth]{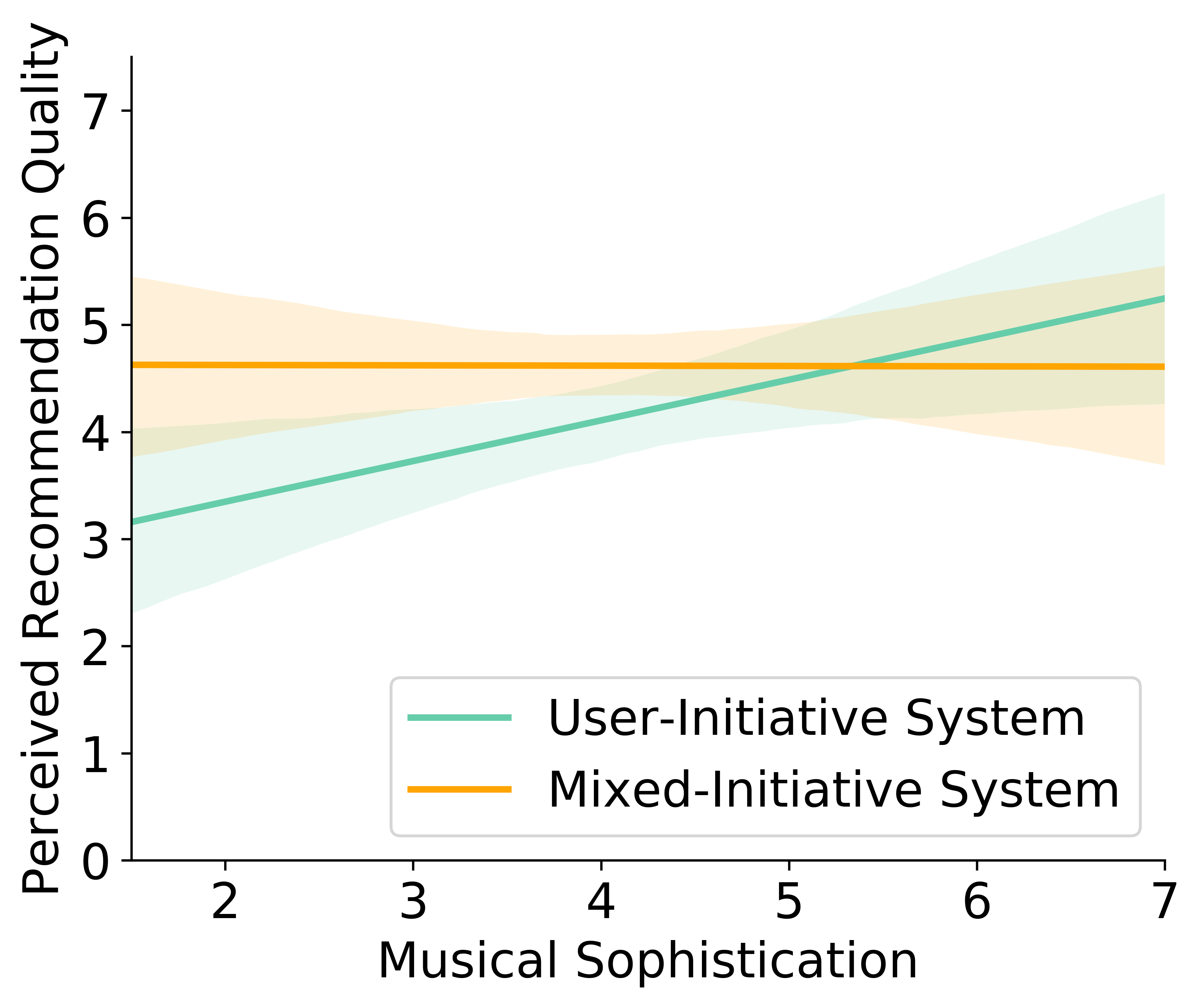} 
    \label{fig:interaction-system-4}}
    \vspace{-0.2cm}
    \caption{Interaction effects between personal characteristics and initiative strategy on users' trust-related perception constructs. (a-c) \textit{Conscientiousness} (C): Users with higher C tended to have a better perception and showed more trust in the Mixed-Initiative system. (d) \textit{Musical Sophistication} (MS): Users with higher MS tended to perceive higher recommendation quality from the User-Initiative system.}
    \Description{Figure 4 illustrates the interaction effects between personal characteristics and initiative strategy on users’ trust-related perception constructs. Subfigures (a-c) show that users with higher Conscientiousness tended to have a better perception and showed more trust in the Mixed-Initiative system. Subfigure (d) shows that users with higher Musical Sophistication tended to perceive higher recommendation quality from the User-Initiative system.}
    \label{fig:interaction-effect-system}
    \vspace{-0.3cm}
\end{figure*}

\begin{figure*}[t]
    \vspace{-0.4cm}
    \centering
    \subfigure[Interaction effect of  \textit{conscientiousness} and \textit{task}  \textit{complexity}  on   Perceived  Recommendation Quality.]
    {\includegraphics[width=0.35\textwidth]{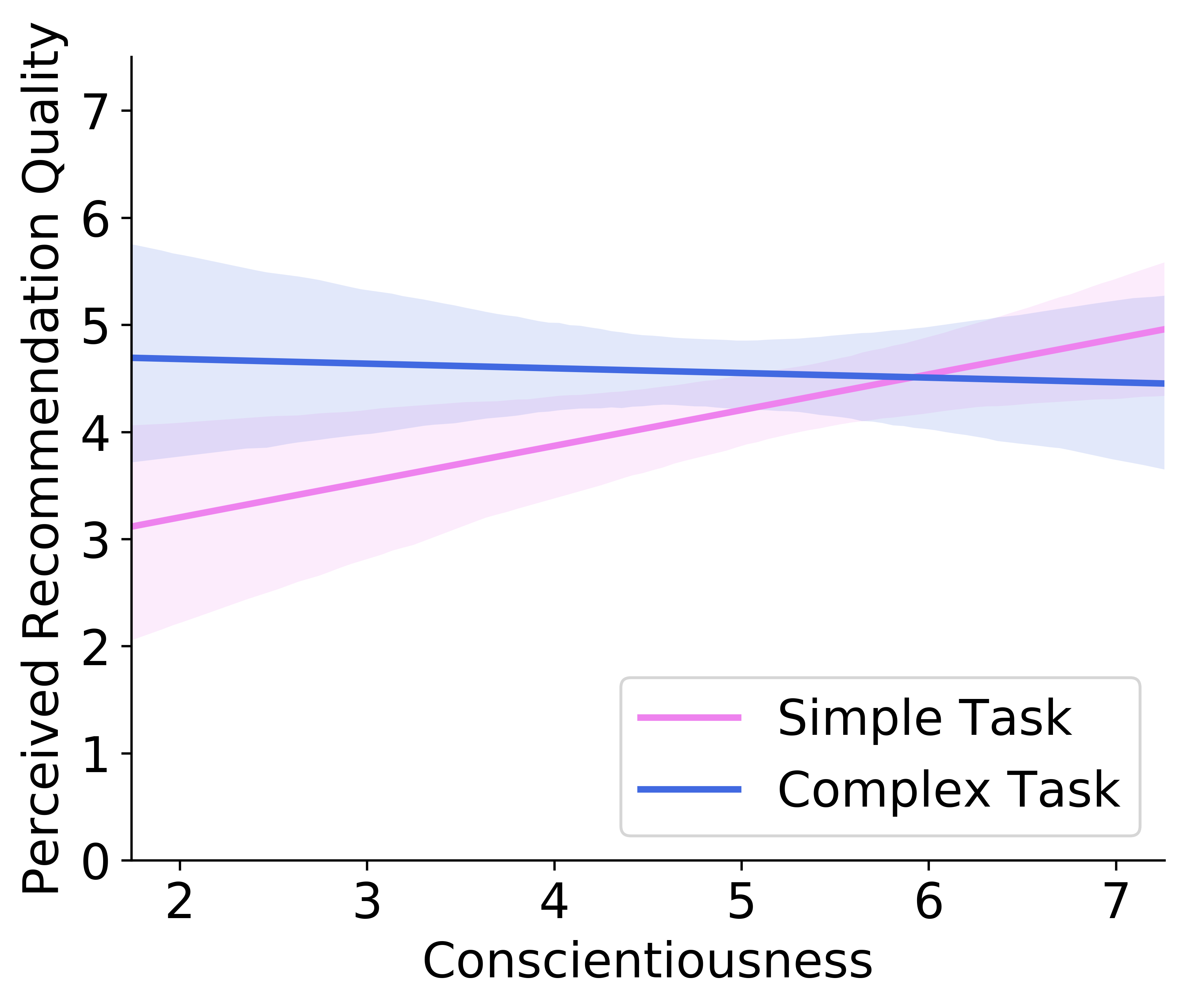} 
    \label{fig:interaction-task-1}}
     \hspace{2cm}
    \subfigure[Interaction effect of \textit{trust}   \textit{propensity} and \textit{task}  \textit{complexity}  on   Perceived Effort.]
    {\includegraphics[width=0.35\textwidth]{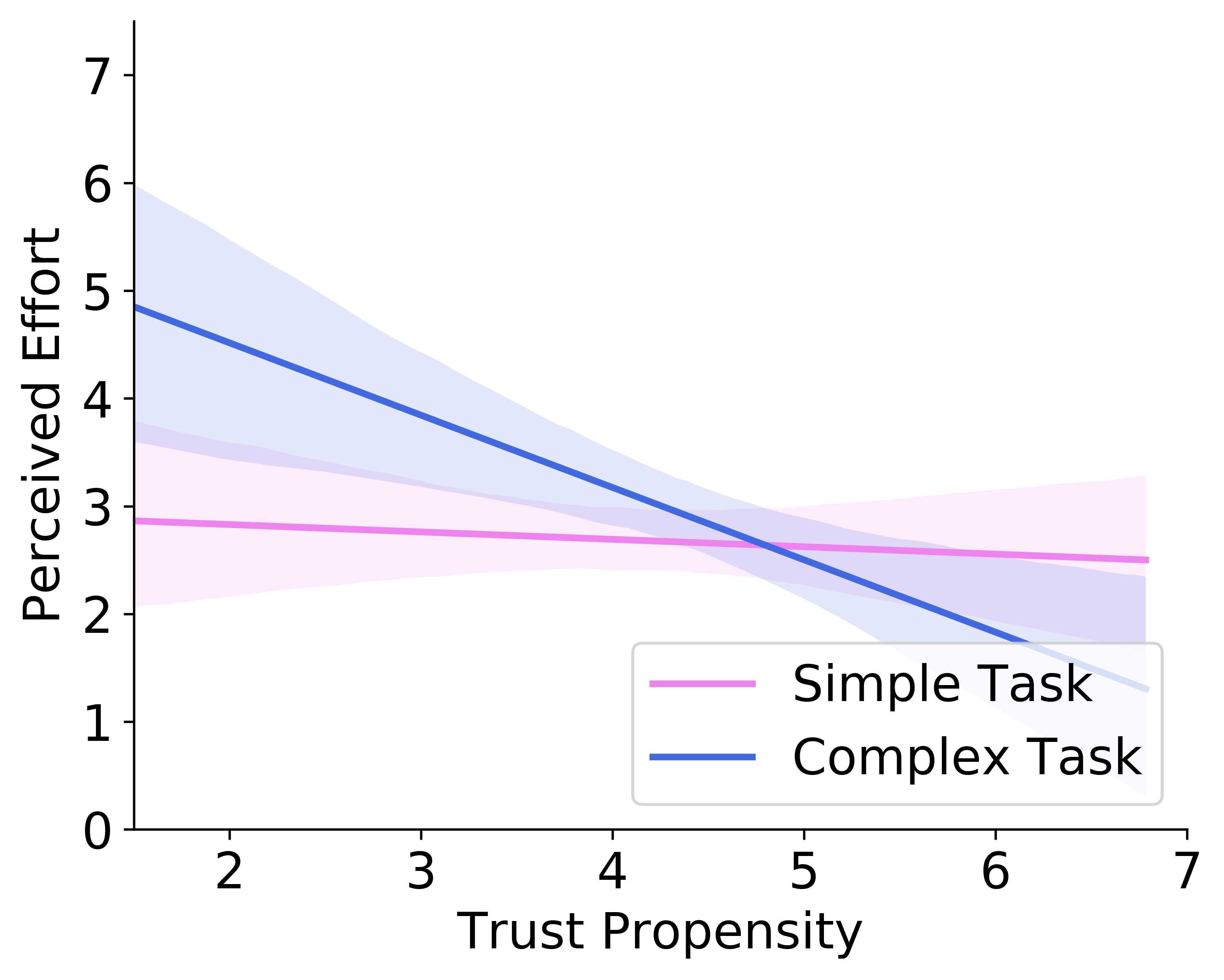} 
    \label{fig:interaction-task-2}}\\ 
    \vspace{-0.3cm}
    \subfigure[Interaction effect of \textit{trust}     \textit{propensity} and \textit{task}  \textit{complexity}  on   Intention to Use.]
    {\includegraphics[width=0.35\textwidth]{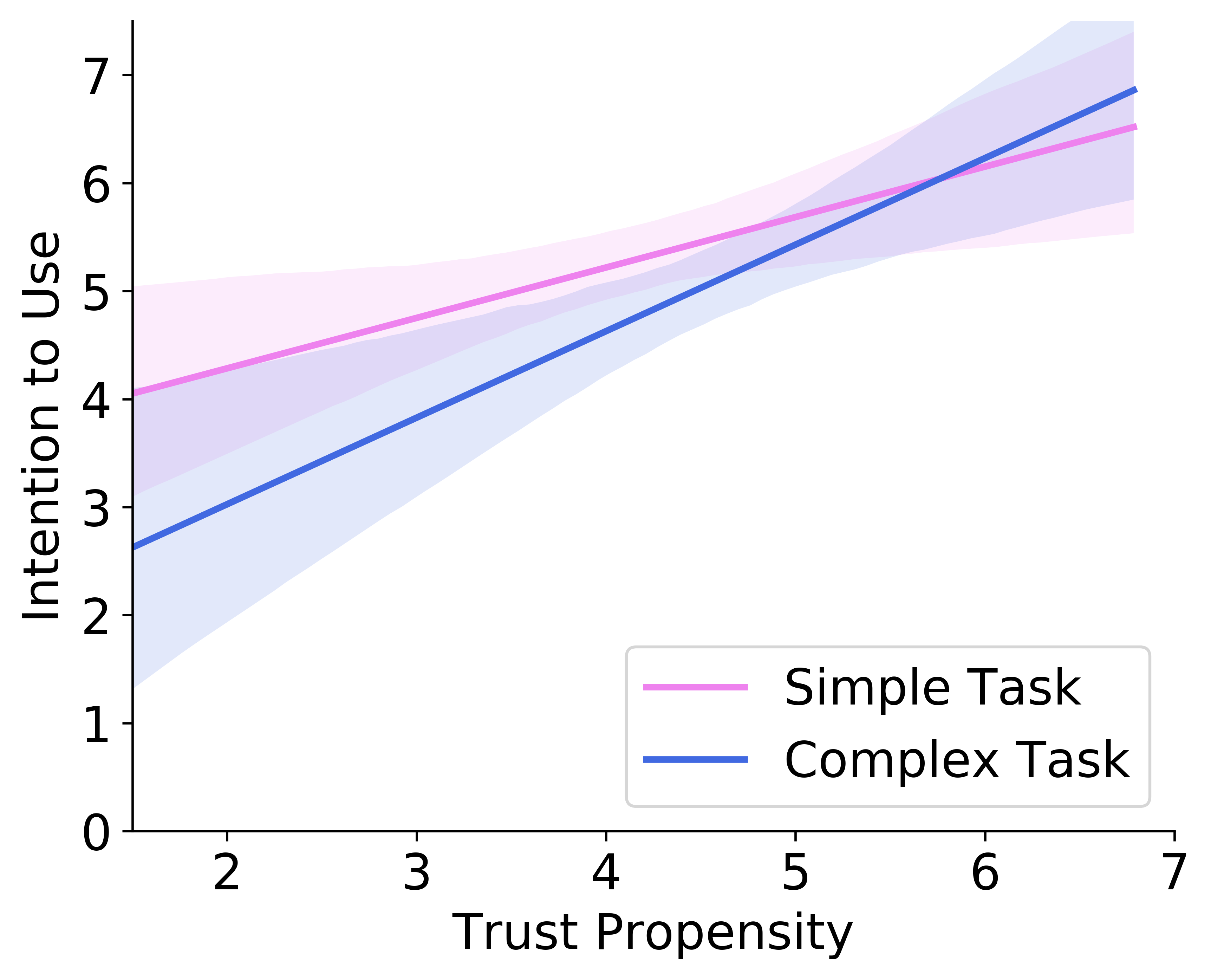}
    \label{fig:interaction-task-3}}
    \hspace{2cm}
    \subfigure[Interaction effect of   \textit{musical}  \textit{sophistication} and \textit{task}  \textit{complexity}  on  Perceived   Recommendation Quality.]
    {\includegraphics[width=0.35\textwidth]{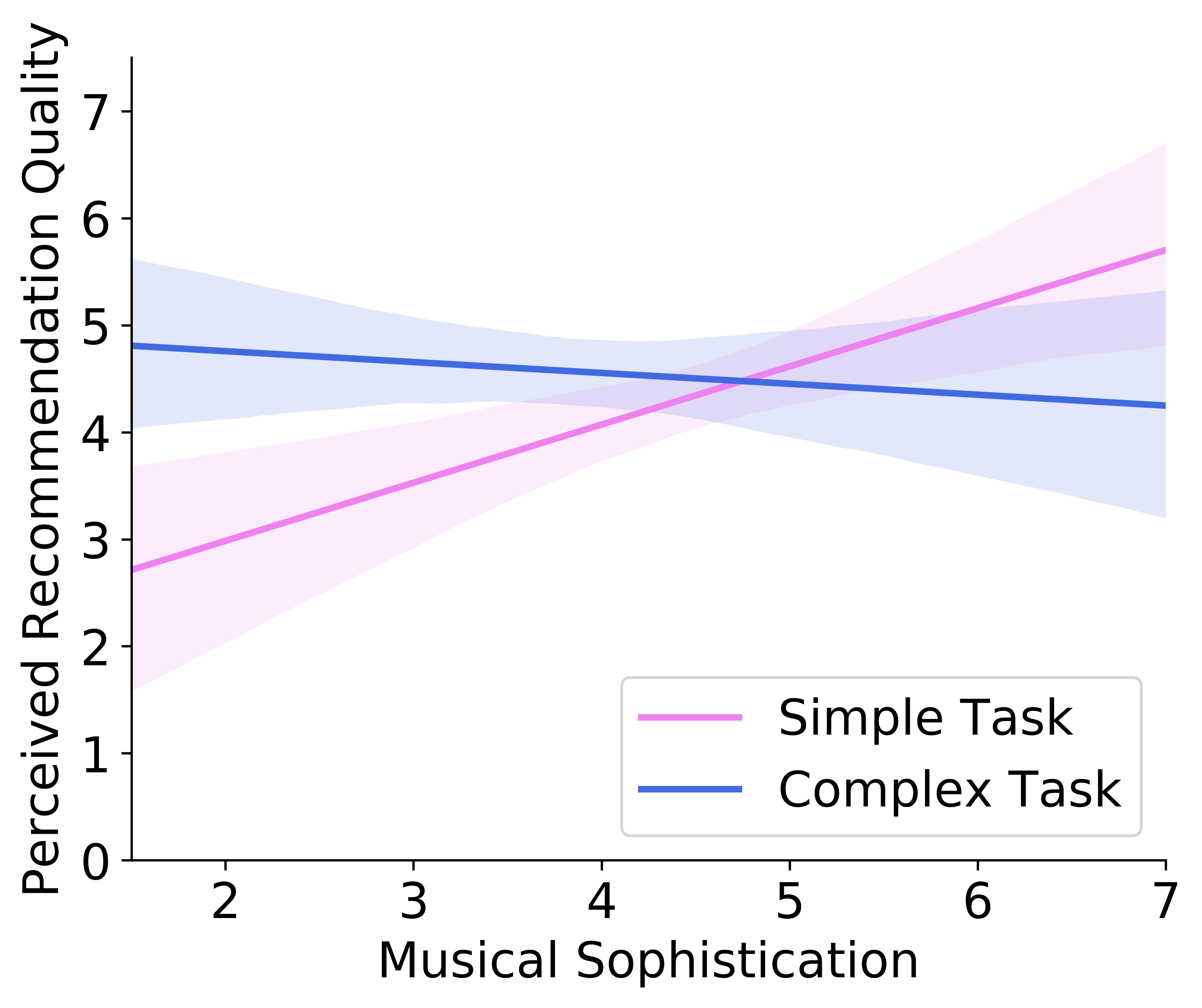} 
    \label{fig:interaction-task-4}}
    \vspace{-0.3cm}
    \caption{Interaction effects between  personal characteristics  and task complexity on users' trust-related perception constructs.  (a) \textit{Conscientiousness} (C): C showed a positive effect on the users' perceived recommendation quality for the Simple Task. (b-c) \textit{Trust Propensity} (TP): The effects of TP on users' trust-related perception were stronger for the Complex Task. (d) \textit{Musical Sophistication} (MS): Users with higher MS tended to have a better perception of recommendations for the Simple Task.}
    \Description{Figure 5 illustrates the interaction effects between personal characteristics and task complexity on users’ trust-related perception constructs. Subfigure (a) shows that Conscientiousness showed a positive effect on the users’ perceived recommendation quality for the Simple Task. Subfigures (b and c) show that the effects of Trust Propensity on users’ trust-related perception were stronger for the Complex Task. Subfigure (d) shows that users with higher Musical Sophistication tended to have a better perception of recommendations for the Simple Task.}
    \label{fig:interaction-effect-task}
    \vspace{-0.3cm}
\end{figure*}

As inspired by previous studies~\cite{Bart:RecSys2011,Myers:CHI2019}, individual users may have different perceptions of the two conversational recommenders (User-Initiative and Mixed-Initiative systems), and may show different attitudes when performing the two user tasks (Simple Task and Complex Task), which may influence their formation of trust in the CRS. Therefore, we investigated how the user-related factors (personal characteristics) interact with the system-related factor (initiative strategy) and the context-related factor (task complexity) to influence user trust in the CRS. Specifically, we used linear regression models to process the mix of numerical and categorical independent variables, namely personal characteristics, initiative strategy and task complexity as the independent variables, and the five trust-related perception constructs (Table~\ref{tab:questionnaire}) as the dependent variables. Table~\ref{tab:regression} presents the results of the regression models that show how users' trust-related perception is influenced by personal characteristics, initiative strategy, and task complexity, revealing their interaction effects (represented by interaction terms in the model). We report coefficients, standard errors, p-values, $R^2$ and adjusted $R^2$ values. 

\subsubsection{Interaction Effects between  Personal Characteristics and Initiative Strategy, Task Complexity}

We detected a significant three-way interaction effect between the trait \textit{agreeableness}, \textit{initiative strategy} and \textit{task complexity} on users' perceived conversational interaction. Specifically, when using the Mixed-Initiative system to accomplish the Complex Task, users' \textit{agreeableness} positively affected their perceptions of the conversation interaction ($r$ = 0.40, $p$ < .05, 95\%  confidence interval [CI]: [0.08, 0.64]).\footnote{Here we conducted Spearman's correlation analyses after detecting interaction effects to 
clearly show the relationship between a personal characteristic and a user perception construct in a particular condition. We followed this procedure to analyze all the detected interaction effects.} In other words, system-initiative suggestions help users explore music, and users with higher \textit{agreeableness} are likely to have a better experience with such conversational interaction. However, no significant correlations were detected in the other three experimental conditions.

\subsubsection{Interaction Effects between  Personal Characteristics and Initiative Strategy}
Table~\ref{tab:regression} shows significant interaction effects between \textit{initiative strategy} and the two personal characteristics, \textit{conscientiousness} and \textit{musical sophistication}: 

\begin{itemize}
    \item \textbf{Conscientiousness}. The models in Table~\ref{tab:regression} show significant interaction effects between the trait \textit{conscientiousness} and \textit{initiative strategy} on several trust-related perception constructs, including perceived recommendation quality, perceived conversational interaction, and perceived trust. Figures~\ref{fig:interaction-system-1}, ~\ref{fig:interaction-system-2} and~\ref{fig:interaction-system-3} visualize these interaction effects. In the Mixed-Initiative system,  users' \textit{conscientiousness} levels positively influenced their perceived recommendation quality  ($r$ = 0.36, $p$ < .001, 95\% CI: [0.15, 0.53]), perceived conversational interaction ($r$ = 0.41, $p$ < .001, 95\% CI: [0.21, 0.57]), and perceived trust ($r$ = 0.39, $p$ < .001, 95\% CI: [0.18, 0.56]). In contrast, in the User-Initiative system, the trait \textit{conscientiousness} was not correlated with users' trust-related perception. Conscientious users may be more cautious and consider more choices when making a decision~\cite{john1999big}, so they may be more inclined to appreciate the suggestions offered by the Mixed-Initiative system that can guide them to discover more music when finding songs of interest.  

    \item \textbf{Musical Sophistication}. As for \textit{domain knowledge}, we detected an interaction effect between \textit{musical sophistication} and \textit{initiative strategy} on users' perceived recommendation quality. As illustrated in Figure~\ref{fig:interaction-system-4}, we can see that users with higher \textit{musical sophistication} tended to have a better perception of recommendations in the User-Initiative system ($r$ = 0.21, $p$ < .1, 95\% CI: [-0.03, 0.43]), whereas in the Mixed-Initiative System, the level of \textit{musical sophistication} did not have a significant influence. We also observed that users of lower \textit{musical sophistication} tended to perceive higher recommendations quality in the Mixed-Initiative system than in the User-Initiative system, implying that the system's suggestions are more helpful for domain novices. 
\end{itemize}

\begin{table*}[t]
\vspace{-0.3cm}
\centering
\footnotesize
\caption{Summary of the major findings. The positive sign (+) and the negative sign (-) indicate significant positive effects and negative effects, respectively} 
\label{tab:summary}
\begin{tabular}{p{0.2cm}lp{10pc}p{10pc}p{10pc}}
\toprule
\multicolumn{2}{l}{} & \multicolumn{1}{c}{} & \multicolumn{1}{c}{\textbf{Interaction Effect with}} & \multicolumn{1}{c}{\textbf{Interaction Effect with}} \\
\multicolumn{2}{l}{\textbf{Personal Characteristic}} & \multicolumn{1}{c}{\textbf{Direct Effect}} & \multicolumn{1}{c}{\textbf{Initiative Strategy}} & \multicolumn{1}{c}{\textbf{Task Complexity}} \\
\midrule
\multicolumn{3}{l}{\textbf{Big-Five Personality Traits}} &  \\   \midrule
 & \textbf{Conscientiousness} & (+): \newline Perceived Conversational Interaction & (+) in Mixed-Initiative:\newline Perceived Recommendation Quality;\newline Perceived Conversational Interaction;\newline Perceived Trust & (+) in Simple Task:\newline Perceived Recommendation Quality \\ \cmidrule(l){2-5} 
 & \textbf{Extroversion} & (+): \newline Perceived Recommendation Quality &  &  \\ \cmidrule(l){2-5} 
 & \textbf{Agreeableness} &  & \multicolumn{2}{l}{(+) in Mixed-Initiative \& Complex Task: Perceived Conversational Interaction} \\ \midrule
\multicolumn{2}{l}{\textbf{Trust Propensity}} & (+): \newline Perceived Conversation Interaction;\newline Intention to Use &  & (-) in Complex Task:\newline Perceived Effort\newline (+) in Complex Task > Simple Task:\newline Intention to Use \\ \midrule
\multicolumn{2}{l}{\textbf{Music Sophistication}} & (+): \newline Intention to Use & (+) in User-Initiative: \newline Perceived Recommendation Quality & (+) in Simple Task: \newline Perceived Recommendation Quality \\
\bottomrule
\end{tabular}
\vspace{-0.3cm}
\end{table*}

\subsubsection{Interaction Effects between Personal Characteristics and Task Complexity}

From Table~\ref{tab:regression}, significant interaction effects were detected between \textit{task complexity} and three personal characteristics, \textit{conscientiousness}, \textit{trust propensity}, and \textit{musical sophistication}: 

\begin{itemize}
    \item \textbf{Conscientiousness}. We found a significant interaction effect between the trait \textit{conscientiousness} and \textit{task complexity} on users' perceived recommendation quality. As visualized in Figure~\ref{fig:interaction-task-1}, the positive effect of \textit{conscientiousness} on perceived recommendation quality was observed when users perform the Simple Task ($r$ = 0.32, $p$ < .01, 95\% CI: [0.10, 0.51]), but no relationship was found for the Complex Task. Together with the results in Figure~\ref{fig:interaction-system-1}, a crossover interaction effect was observed between \textit{conscientiousness} and \textit{initiative strategy}, suggesting that when users perform the Complex Task, their \textit{conscientiousness} levels may differently influence their perceived recommendation quality, depending on the system's initiative strategy (user-initiative or mixed-initiative).

    \item \textbf{Trust Propensity}. \textit{Task complexity} influenced the effects of \textit{trust propensity} on users' perceived effort and intention to use the conversational recommender. Specifically, users with higher \textit{trust propensity} levels tended to feel less effort using the conversational recommender to perform the Complex Task ($r$ = -0.34, $p$ < .01, 95\% CI: [-0.53, -0.12]), but the correlation between them was not obvious regarding the Simple Task [see Figure~\ref{fig:interaction-task-2}], probably due to the intrinsically lower user effort required for the Simple Task. Moreover, \textit{trust propensity} positively influenced users' intention to use the conversational recommender (also shown in Figure~\ref{fig:SEM-Trust}), and Figure~\ref{fig:interaction-task-3} shows that the positive effect was stronger when users performed the Complex Task ($r$ = 0.32, $p$ < .01, 95\% CI: [0.09, 0.52]) than the Simple Task ($r$ = 0.27, $p$ < .05, 95\% CI: [0.04, 0.46]).

    \item \textbf{Musical Sophistication}. A significant interaction effect was detected between \textit{musical sophistication} and \textit{task complexity} on users' perceived recommendation quality. As shown in Figure~\ref{fig:interaction-task-4}, when performing the Simple Task, users with higher \textit{musical sophistication} tended to have a more positive perception of recommendations than users with lower \textit{musical sophistication} ($r$ = 0.34, $p$ < .01, 95\% CI: [0.12, 0.52]), which could be due to the higher skill levels of music professionals for tuning recommendations to find songs that suit their tastes. 
\end{itemize}

Table~\ref{tab:summary} summarizes the effects of the three personal characteristics on user trust toward the conversational music recommenders and their interaction effects with the initiative strategy (User-Initiative and Mixed-Initiative) and with the task complexity (Simple Task and Complex Task). Overall, \textit{trust propensity} and \textit{musical sophistication} directly influenced users' intention to use, and \textit{conscientiousness} interacted with the initiative strategy to influence users' perceived trust in the CRS.

\section{Discussion and Design Implications}
In this research, we have sought to better understand user trust in conversational recommender systems (CRSs). By examining the relationships between users' perceptions of system competence (especially recommendation quality and conversational interaction) and their trust, we found that users' experience with conversational interaction was particularly important for inspiring user trust toward the conversational recommender (high $\beta$ coefficients for the significant paths, as shown in Figure~\ref{fig:SEM-Trust}). As driven by the three-layered trust model~\cite{hoff2015trust}, we investigated the influences of three types of factors (user-related, system-related, and context-related) on user trust in CRSs, in which we highlight the impacts of user-related factors (users' Big-Five \textit{personality traits}, \textit{trust propensity}, and \textit{domain knowledge}). This section will discuss the key findings of our study and their implications for designing trustworthy CRSs.

\subsection{Key Findings}
\textbf{Key Finding \#1: Users with higher \textit{conscientiousness} have a better perception of system competence and show more trust toward the Mixed-Initiative system.}
Our results demonstrate that users with a higher level of \textit{conscientiousness} have more positive perceptions in terms of both recommendations and conversational interaction with the Mixed-Initiative system, engendering higher trust in the CRS [see Figures~\ref{fig:interaction-system-1}, ~\ref{fig:interaction-system-2} and~\ref{fig:interaction-system-3}]. This finding is in line with previous studies showing that more conscientious people have higher trust in automation when conducting decision-making tasks~\cite{chien2016relation, cho2016effect}. Highly conscientious users tend to be cautious, responsible~\cite{john1999big}, and may have maximising tendencies (i.e., the tendency to explore and compare alternatives, and look for the best option)~\cite{miceli2018personality}, which may result in more appreciation for the suggestions from the system that may help them become more informed to make a confident decision. This finding also suggests that individual differences in users' decision-making style, i.e., maximizing (examining more alternatives to select the best option) and satisficing (settling for a good-enough option)~\cite{schwartz2002maximizing, Michael:UMAP2018}, may be influential on user trust in CRSs, which can be investigated in future research.

\textit{Design Implications:}
Trustworthy CRS design should consider users' personality traits, especially \textit{conscientiousness}. For users with higher \textit{conscientiousness} who like to carefully consider all facets before making a choice, the Mixed-Initiative system that supports both user-initiative and system-initiative interactions is more desirable. System-initiated guidance may support conscientious users in seeking alternatives and finding the  ``perfect'' items from recommendations, hence fostering user trust toward the system. However, for users with lower \textit{conscientiousness}, the level of system-initiative can be relatively lower because those users tend to be casual and impulsive and might not appreciate extensive guidance from the system.

\textbf{Key Finding \#2: Users' \textit{trust propensity} positively influences user trust in conversational recommenders, but the degree of influence is affected by the task complexity.}  
Our results imply the positive effects of \textit{trust propensity} on users' perceptions of the conversational interaction and their intention to use, which is consistent with previous reports of the positive effect of one's general tendency to trust others or technology on trust in recommender systems~\cite{chen2005trust, Wang:JMIS2007}. Moreover, the complexity of the performed task tends to strengthen this effect [see Figures~\ref{fig:interaction-task-2} and~\ref{fig:interaction-task-3}], suggesting a stronger influence of \textit{trust propensity} when users perform the Complex Task. We found that users with higher \textit{trust propensity} perceived much less effort and higher intention to use the system than users with lower \textit{trust propensity}, but this trend was more significant for the Complex Task than the Simple Task.  We argue that, when performing a complex task, users with higher \textit{trust propensity} are more likely to take advantage of an effective conversational interaction to indicate what they like or dislike and obtain system guidance when they get stuck on a task. However, as shown in our model (Figure~\ref{fig:SEM-Trust}), users with lower \textit{trust propensity} benefit less from conversational interaction, which has a strong influence on user trust (in terms of both perceived conversational interaction and intention to use).

\textit{Design Implications:}
CRS researchers have attempted to improve recommendation quality and conversation interaction to build user trust in the system. However, previous studies have not adapted the design of trustworthy CRSs to users' \textit{trust propensity}. The ``one size fits all'' approach can be flawed because it assumes all users have the same \textit{trust propensity} level. Thus, future design of CRSs could also consider users' general tendency to trust technology. For example, the system may help users with lower \textit{trust propensity} understand more about the system's ability and guide them to accomplish simple tasks in the initial period, which would improve their initial trust in the system's competence.

\textbf{Key Finding \#3: Users with stronger \textit{domain knowledge} have a higher intention to use conversational recommenders and prefer to explore recommendations by themselves.}
Our results indicate that users with more \textit{domain knowledge} (i.e.,  higher \textit{musical sophistication} in our case) have a higher intention to use the CRS. Furthermore, users with a higher level of \textit{domain knowledge} benefit more from the conversational interaction with recommendations, because they possess a greater ability to articulate their preferences than do domain novices~\cite{jin2018effects}. 
In addition, system-initiative suggestions are more helpful for users with less \textit{domain knowledge} when looking for recommendations. In contrast, domain-knowledgeable users tended to have a better perception in finding recommendations by themselves, probably because this type of user desires more control over their decisions~\cite{Bart:RecSys2011}.  

\textit{Design Implications:} 
This finding informs that the users' \textit{domain knowledge} level should be taken into account in the design of CRSs, because it influences users' intention to use the system as well as their preferred initiative strategies.
For example, the Mixed-Initiative system is more beneficial for novice users as they may need more suggestions from the system to find recommendations that fit their interests. In contrast, the User-Initiative system might be sufficient for domain experts because they often expect higher control over the interaction with the system and to be interrupted less by the system-initiated guidance.

\subsection{Limitations}
Before concluding this paper, we highlight some limitations of our research. First, the factors that influence user trust in conversational systems are not limited to Competence Perception, which was the only dimension investigated in our study. Anthropomorphism~\cite{seeger2018human}, security and privacy~\cite{folstad2018makes} are additional relevant dimensions of user trust. However, these dimensions are frequently discussed in the context of user trust in customer service chatbots and are influenced by additional personal characteristics, such as affective states~\cite{airenti2018development} and privacy concerns\cite{saglam2021privacy}. To avoid added complexity, our trust model mainly considers the dimension of Competence Perception of CRSs, namely, perceived recommendation quality, perceived effort, and perceived conversational interaction. Second,
recommender systems are applied in various domains including media, e-commerce, and healthcare. However, we conducted our study with a CRS designed only for music recommendations, which may limit the generalizability of our findings to other domains. In light of differences in user involvement levels~\cite{Chen:umuai2012}, user trust is more crucial in certain domains, such as e-commerce and healthcare. Future work will validate our findings in different CRS application domains. Third, we only considered a text-based CRS for this investigation, and the results may differ when users interact with a voice-based CRS. In future work, we plan to investigate whether our results are applicable to the voice-based CRS. 



\section{Conclusions}
This study investigated the effects of the three types of factors (user-related, system-related and context-related) on user trust, grounded on the framework of Hoff and Bashir's three-layered trust model~\cite{hoff2015trust}. Our study demonstrated the main effects of user-related factors (personal characteristics) and their interaction effects with the system-related factor (initiative strategy) and the context-related factor (task complexity) on user trust in conversational recommender systems (CRSs). Our findings indicate that \textit{trust propensity} and \textit{domain knowledge} directly influence user trust. Moreover, personal characteristics, like \textit{conscientiousness} and \textit{domain knowledge}, can exert influences on user trust in CRSs with different initiative strategies (\textit{user-initiative} and \textit{mixed-initiative}). 


Prior work on user trust toward traditional recommender systems~\cite{chen2005trust, Shlomo:IUI2017} has highlighted the significance of measuring competence perception based on recommendation quality, whereas we emphasize the importance of gauging perceived conversational interaction because it has a stronger influence on user trust in CRSs. As the initiative strategy influences the way users interact with the CRS, we also highlight the interaction effects of personal characteristics and initiative strategy on user trust. Our findings contribute to the research community of Human-AI interactions~\cite{amershi2019guidelines} and will be of interest to researchers who investigate the role of personalization in building user trust in conversational AI systems and the impacts of personal characteristics when developing trustworthy AI systems such as CRSs.

\begin{acks}
The work was supported by Hong Kong Research Grants Council (RGC/HKBU12201620) and Hong Kong Baptist University IRCMS Project (IRCMS/19-20/D05).  We also thank all participants for
their time in taking part in our experiment and reviewers for their
constructive comments on our paper.
\end{acks}
\balance
\bibliographystyle{ACM-Reference-Format}
\bibliography{paper}

\end{document}